\documentclass[twocolumn,showpacs,aps]{revtex4-2}
\usepackage[utf8]{inputenc}
\usepackage{graphicx}
\usepackage{longtable}
\usepackage{subfigure}
\usepackage[tbtags]{amsmath}
\usepackage{booktabs}
\usepackage{makecell}
\usepackage{diagbox}
\usepackage{lastpage}
\usepackage{amssymb,bm,mathrsfs,bbm,amscd}
\usepackage{wrapfig}
\usepackage{dcolumn,booktabs}
\usepackage{braket}
\usepackage{ifpdf}
\usepackage{cancel}

\newcolumntype{d}[1]{D{.}{.}{#1}}
\usepackage[export]{adjustbox}

\usepackage{color}

\renewcommand{\u}{\uparrow}
\renewcommand{\d}{\downarrow}

\begin{document}

\title{Ferromagnetic domains in the large-$U$ Hubbard model with a few holes: an FCIQMC study}

\author{Sujun Yun}
\email{yunsujun@163.com}
\affiliation{Max Planck Institute for Solid State Research, Heisenbergstr. 1, 70569 Stuttgart, Germany}
\affiliation{School of electronic engineering, Nanjing XiaoZhuang University, Hongjing Road, Nanjing 211171, China}

\author{Werner Dobrautz}
\email{dobrautz@chalmers.se}
\affiliation{Department of Chemistry and Chemical Engineering,
Chalmers University of Technology, 41296 Gothenburg, Sweden}

\author{Hongjun Luo}
\email{h.luo@fkf.mpg.de}
\affiliation{Max Planck Institute for Solid State Research, Heisenbergstr. 1, 70569 Stuttgart, Germany}

\author{Vamshi Katukuri}
\affiliation{Max Planck Institute for Solid State Research, Heisenbergstr. 1, 70569 Stuttgart, Germany}%

\author{Niklas Liebermann}
\affiliation{Max Planck Institute for Solid State Research, Heisenbergstr. 1, 70569 Stuttgart, Germany}%

\author{Ali Alavi}
\email{a.alavi@fkf.mpg.de}
\affiliation{Max Planck Institute for Solid State Research, Heisenbergstr. 1, 70569 Stuttgart, Germany}%
\affiliation{Yusuf Hamied Department of Chemistry, University of Cambridge, Lensfield Road, Cambridge CB2 1EW, United Kingdom}%

\begin{abstract}
Two-dimensional Hubbard lattices with two or three holes are investigated as a function of $U$ in the large-$U$ limit. In the so-called Nagaoka limit (one-hole system at infinite $U$), it is known that the Hubbard model exhibits a ferromagnetic ground-state. Here, by means of exact FCIQMC simulations applied to periodic lattices up to 24 sites, we compute spin-spin correlation functions as a function of increasing $U$. The correlation functions clearly demonstrate the onset of ferromagnetic domains, centred on individual holes. The overall total spin of the wavefunctions remain the lowest possible (0, or $\frac{1}{2}$, depending on the number of holes). The ferromagnetic domains appear at interaction strengths comparable to the critical interaction strengths of the Nagaoka transition in finite systems with strictly one hole.  The existence of such ferromagnetic domains is the signature of Nagaoka physics in Hubbard systems with a small (but greater than one) number of holes.

\end{abstract}
\maketitle
\section{Introduction}


The Hubbard model is a simple yet important model in the study of correlated electrons, as it captures complex correlations between electrons on a lattice with a fairly simple Hamiltonian~\cite{Qin2022}. Exact results of two-dimensional Hubbard model are helpful for understanding a plethora of phenomena in strongly correlated systems, including
pairing mechanisms in unconventional superconductors~\citep{scalapino2012}, the Mott metal-insulator transition~\cite{imada1998}, optical conductivity~\cite{Tohyama2005,Daigo2018}, and itinerant magnetism~\cite{dai2022,Makus2020}. For the single-band 2D Hubbard model on a square lattice,
Nagaoka~\cite{Nagaoka1966} proved analytically that the limit of infinitely strong interactions, in the presence of a single hole on top of a Mott-insulating state with one electron per site, results in a ferromagnetic ground state. Intuitively, the Nagaoka ferromagnetism can be understood as resulting from an interference effect between the different paths that the hole can take across the lattice. When the spins are aligned, these paths interfere constructively and lower the kinetic energy of the hole~\cite{Thouless1965,lib1989,Tasaki1989,Tasaki1998}.

 
 While Nagaoka ferromagnetism has been analytically proven under extreme conditions, and has also been observed in a quantum dot plaquette~\cite{Ivantsov2017}, the stability of the Nagaoka ferromagnetic state at finite interaction strengths on finite lattices has also been actively studied ~\cite{Barbieri1990,Becca2001,Liu2012,Hartmann1993,Hartmann1996,wen1989,fan1989,yun2021}. However, open questions still exist, especially concerning the thermodynamic stability of the ferromagnetic state for systems with more than one hole. Extrapolations from the results on finite lattices have been used to study properties in the thermodynamic limit. Thus it is important to obtain exact results in systems with two-, three-, and perhaps more-hole systems, on lattices as large as possible.

In the two-hole system, the total spin of the ground-state is zero ($S=0$). This has been numerically studied by exact diagonalization (ED)~\cite{riera1989},   the spin-adapted full configuration interaction quantum Monte Carlo (FCIQMC) method~\cite{yun2021} and analytical studies of arbitrarily large systems~\cite{wen1989}. However, the specific type of magnetism is unknown, because anti-ferromagnetism, para-magnetism, as well as low-spin-coupled ferromagnetic domains,  all correspond to $S=0$. 
For the three-hole system, on the other hand, ED results of an effective Hamiltonian~\cite{riera1989} show that the total spin of ground-states on the 8- and 16-site lattices are $\frac{3}{2}$ and $\frac{7}{2}$ respectively. Here the effective Hamiltonian is constructed to exclude double occupation and thus should be related to the Hubbard model in the large $U$ limit.
It is interesting to see whether the partial magnetization will still be observed on larger lattices. 
In order to answer the above open questions, in this work we investigate the two- and three-hole systems with the full configuration interaction quantum Monte Carlo (FCIQMC) method.

FCIQMC is based on stochastic simulations of the dynamic evolution of the wave function in imaginary time.
Different to the diffusion quantum Monte Carlo (DMC)~\cite{Foulkes2001} with the fixed node approximation and  the auxiliary-field quantum Monte Carlo (AFQMC)~\cite{Shi2021} with the phaseless approximation, no systematic approximation is made in FCIQMC~\cite{Booth2009, Cleland_2010, Guther2020}, and it thus serves as a highly accurate method to approach the ground state wave functions. The annihilation procedure of the algorithm enables it to overcome the fermionic sign problem exactly, as long as enough walkers are used. In practice, for Hubbard systems at large $U$, this means, with the currently available hardware, systems up to 26 sites can be studied~\cite{yun2021}. (A 26-site lattice represents a useful increase in size compared to exact-diagonalisation, for which 20-sites is the largest lattice size so far reported~\cite{Tohyama2005}). In the present paper, we extend this to the study of a system with a few holes, as well as report on spin-spin correlation functions for the obtained exact ground-states. 

The software NECI, a state-of-the-art implementation of the FCIQMC algorithm, utilizes a very powerful parallelization and scales efficiently to more than 24000 central processing unit cores~\cite{Guther2020}. 
The FCIQMC method in a Slater determinant (SD) basis has been extended to calculate ground and excited state energies, spectral and Green’s functions for \textit{ab initio} and model systems, as well as properties via the one-, two-, three- and four-body reduced density matrices(RDMs). To study magnetism, we need to use the replica-sampled 2-RDMs~\cite{Overy2014,Thomas2015,Blunt2017}
to obtain the spatial spin distribution. The replica-sampling technique removes the systematic error in the RDM, at the expense of requiring a second walker distribution. The premise is to ensure that these two walker distributions are entirely independent and propagated in parallel, sampling the same (in this
instance ground-state) distribution. This ensures an unbiased sampling of the desired RDM, by ensuring that each RDM contribution is derived from the product of an uncorrelated amplitude from each
replica walker distribution. 
By using replica-sampled 2-RDMs \, 
the spin-spin correlation function, $\langle\hat{\bf S}_i\cdot \hat{\bf S}_j\rangle$,  can be calculated, where $i$ and $j$ are
lattice site indices. This spin-spin correlation function can then be used to identify the specific type of magnetism of the ground states.  


FCIQMC in a spin-adapted basis is also used to study the partial polarization in three-hole systems. Spin-adapted FCIQMC uses $SU(2)$ symmetry (arising from the vanishing commutator $[\hat{H},\hat{{\bf S}}^2]=0$) conservation.
$SU(2)$ symmetry is imposed via the graphical unitary group approach (GUGA)~\cite{Paldus1976,Shavitt1977,Shavitt1978} which dynamically constrains the total spin $S$ of a multi-configuration and highly open-shell wave function in an efficacious manner. The spin-adapted version of the FCIQMC algorithm based on GUGA has been developed in our group~\cite{Wernerandsimon2019, Dobrautz2021}, with -- among others -- applications to \emph{ab initio} system~\cite{Li_Manni_2020, giovanni2021} and Nagaoka ferromagnetism in one-hole system~\cite{yun2021}. With the spin-adapted method, the magnetisation of the ground state can be determined in a reliable way, especially for systems with small spin gaps. The results of spin-adapted FCIQMC show the partial spin-polarization only appears in small, three-hole system (less than 18 sites)~\cite{riera1989}, which is the second important result of this work.

The rest of this paper is organized as follows: In Sec. II, we briefly describe the methods, where we mainly provide some more details on the measurements of the spin-spin correlation function, $\langle \hat{\bf S}_i\cdot\hat{\bf S}_j\rangle$ from the replica-sampled 2-RDMs in FCIQMC. In Sec. III, results about the spatial spin distribution and partial spin-polarization are discussed. Finally, we conclude in Sec. IV.


\section{methods}

The Hamiltonian of the Hubbard model in real space takes the form
\begin{equation}
\hat{H} = - t \sum_{\langle i j \rangle \sigma}a^{\dagger}_{i, \sigma} a_{j, \sigma} + U \sum_{i} n_{i\uparrow} n_{i\downarrow} 
\label{oriHamil}
\end{equation}
where $a^{\dagger}_{i \sigma}$ ($a_{i \sigma}$) creates (annihilates) an electron with spin
$\sigma$ on site $i$, and $ n_{i\sigma}=a^{\dagger}_{i \sigma}a_{i \sigma}$ is the particle number operator.
$U$ refers to the Coulomb interaction strength. We consider only nearest neighbour hopping terms, where $t$ is positive and is used as the unit of the energy. 
When $U$ is infinitely large, there will be no double occupancy and the system can  be treated with an effective Hamiltonian with constrained hopping terms~\cite{riera1989}
\begin{equation}
    H_{\textnormal{eff}}=-t \sum_{\langle i j \rangle \sigma} \tilde{a}^{\dagger}_{i, \sigma} \tilde{a}_{j, \sigma}, 
\label{effHamil}
\end{equation}
with $\tilde{a}^{\dagger}_{i, \sigma}=a_{i, \sigma}(1-n_{i, \sigma})$. 
In our current work, we want to study the magnetic properties for finite $U$ and thus will stay with the original Hamiltonian (\ref{oriHamil}).
Tough, we find that our results for the three-hole systems in the large $U$ limit (see Sec. III(B)), coincide with the result of Riera \emph{et al.}~\cite{riera1989} for the effective Hamiltonian, Eq.~\eqref{effHamil}.
In our investigation we apply two different FCIQMC methods, which are based on full CI expansions in terms of SDs and in terms of spin eigenfunctions (spin-adapted basis states) respectively.

FCIQMC is a projector QMC method for obtaining the ground state wave function $\ket{\Psi_0}$. By Monte Carlo simulation of the imaginary-time evolution of the wave function
\begin{equation}
\ket{\Psi (\tau)} = e^{-\tau (\hat{H}-E_0 )}\ket{\Psi(0)},
\label{eq_Psi_t}
\end{equation}
the ground state wave function is approached in the long time limit $\ket{\Psi(\tau \rightarrow \infty)} \propto \ket{\Psi_0}$. 

In a previous work~\cite{yun2021}, we have investigated the magnetism for one hole and two holes systems by using the spin-adapted ($SU(2)$ conserving) FCIQMC method. We extend these investigations to three-hole systems in this work.
With the spin-adapted method, the magnetisation of the ground state can be determined in a reliable way, especially for systems with small spin gaps. 
For details of the spin-adapted FCIQMC method, we refer to previous work~\cite{yun2021}.

We also calculate the spin-spin correlation function to get a knowledge of the spatial spin distribution. For these calculations, we find that the FCIQMC method based on SDs is more efficient, in particular when it is combined with the replica-sampling techniques for the 2-RDMs. With this technique, two independent FCIQMC simulations are performed in parallel and the wave function expansions 
\begin{equation}
\ket{\Psi(\tau)^{I/II}} = \sum_\mu c_\mu^{I/II}(\tau) |D_\mu\rangle,
\end{equation}
are sampled simultaneously for these two replicas. The sampled coefficients $\{c_{\mu}^I\}$ and  $\{c_{\mu}^{II}\}$ are then used to calculate the 2-RDM elements
\begin{eqnarray}
\Gamma_{pq}^{rs} &=& \langle \Psi|a^{\dagger}_{p}a^{\dagger}_{q} a_{s}a_{r}|\Psi\rangle  \nonumber\\
&=& \sum_{\mu,\nu} c^{I}_\mu c^{II}_\nu \langle D_{\mu}| a^{\dagger}_{p}a^{\dagger}_{q} a_{s}a_{r}|D_{\nu}\rangle,
\label{replica 2-RDM}
\end{eqnarray}
where $p$, $q$, $r$, and $s$ are spin-orbital indices.
Because the two simulations are uncorrelated, the two-body reduced density matrix in Eq.~\eqref{replica 2-RDM} becomes an unbiased one.

To calculate the spin-spin correlation, we need only the following 2-RDM elements 
\begin{eqnarray}
\Gamma_{i\uparrow(\downarrow)j\uparrow(\downarrow)}^{i\uparrow(\downarrow)j\uparrow(\downarrow)} 
 \nonumber & =& \langle \Psi|a^{\dagger}_{i,\uparrow(\downarrow)}a^{\dagger}_{j,\uparrow(\downarrow)} a_{j,\uparrow(\downarrow)}a_{i,\uparrow(\downarrow)}|\Psi\rangle \\
 & =& \langle \Psi | n_{i,\uparrow (\downarrow)} n_{j,\uparrow(\downarrow)} |\Psi \rangle 
\nonumber   \\  
  \Gamma_{i\uparrow(\downarrow)j\downarrow(\uparrow)}^{i\uparrow(\downarrow)j\downarrow(\uparrow)}
& =& \langle \Psi|a^{\dagger}_{i,\uparrow(\downarrow)}a^{\dagger}_{j,\downarrow(\uparrow)} a_{j,\downarrow(\uparrow)}a_{i,\uparrow(\downarrow)}|\Psi\rangle   \nonumber   \\
&  =& \langle \Psi | n_{i,\uparrow (\downarrow)} n_{j,\downarrow(\uparrow)} |\Psi \rangle 
  \nonumber   \\ 
   \Gamma_{i\uparrow(\downarrow)j\downarrow(\uparrow)}^{i\downarrow(\uparrow)j\uparrow(\downarrow)}
 &=& \langle \Psi|a^{\dagger}_{i,\uparrow(\downarrow)}a^{\dagger}_{j,\downarrow(\uparrow)} a_{j,\uparrow(\downarrow)}a_{i,\downarrow(\uparrow)}|\Psi\rangle    .  
  \label{2-RDMsaboutspin}    
\end{eqnarray}
By using the following expressions of the local spin operators
\begin{align}
	S_i^x &= \frac{1}{2}\left( a_{i\u}^\dagger a_{i\d} + a_{i\d}^\dagger a_{i\u} \right) \nonumber \\
	S_i^y &= \frac{i}{2}\left( a_{i\d}^\dagger a_{i\u} - a_{i\u}^\dagger a_{i\d} \right) \nonumber \\
	S_i^z &= \frac{1}{2}\left( n_{i\u} - n_{i\d} \right),
\end{align}
the $\hat S^z$-spin correlation $\langle \hat S_i^z \cdot \hat S_j^z\rangle$ can be evaluated as
 \begin{align}
     \langle &\hat S_i^z \cdot \hat S_j^z\rangle 
 = \frac{1}{4}(\langle \Psi | n_{i,\uparrow} n_{j,\uparrow} |\Psi \rangle- \langle \Psi | n_{i,\uparrow}n_{j,\downarrow} |\Psi \rangle
 \nonumber   \\ 
&-\langle \Psi | n_{i,\downarrow} n_{j,\uparrow} |\Psi \rangle+ \langle \Psi | n_{i,\downarrow} n_{j,\downarrow} |\Psi \rangle )  \nonumber  \\
  &=  \frac{1}{4}(\Gamma_{i\uparrow j\uparrow}^{i\uparrow j\uparrow}+\Gamma_{i\downarrow j\downarrow}^{i\downarrow j\downarrow}    
-\Gamma_{i\uparrow j\downarrow}^{i\uparrow j\downarrow}-\Gamma_{i\downarrow  j\uparrow}^{i\downarrow j\uparrow}),
\label{spin spin correlation along z direction}    
 \end{align}
 and the total spin correlation function as
    \begin{align}
    \langle \hat{\bf S}_i\cdot \hat{\bf S}_j\rangle &= \frac{1}{2}(\Gamma_{i\uparrow j\downarrow}^{i\downarrow j\uparrow}+\Gamma_{i\downarrow  j\uparrow}^{i\uparrow j\downarrow})  \nonumber \\
    +& \frac{1}{4}(\Gamma_{i\uparrow j\uparrow}^{i\uparrow j\uparrow}+\Gamma_{i\downarrow j\downarrow}^{i\downarrow j\downarrow}    
-\Gamma_{i\uparrow j\downarrow}^{i\uparrow j\downarrow}-\Gamma_{i\downarrow  j\uparrow}^{i\downarrow j\uparrow})  
    \end{align}
   
For the special case when $\langle \hat{S}_{z}\rangle = 0$, the above expressions can be simplified to~\cite{Kutzelnigg2010}
\begin{eqnarray}
 \langle \hat{S}^z_i\cdot \hat{S}^z_j\rangle = \frac{1}{2}\left(\Gamma_{i\uparrow j\uparrow}^{i\uparrow j\uparrow} - \Gamma_{i\uparrow j\downarrow}^{i\uparrow j\downarrow}\right) 
 \label{spin spin correlation along z}
 \end{eqnarray}
 and 
 \begin{eqnarray}
 \langle \hat{\bf S}_i \cdot \hat{\bf S}_j\rangle = \Gamma_{i\uparrow j\downarrow}^{i\downarrow j\uparrow} +  \frac{1}{2}(\Gamma_{i\uparrow j\uparrow}^{i\uparrow j\uparrow} - \Gamma_{i\uparrow j\downarrow}^{i\uparrow j\downarrow})
\label{spin spin correlation for close shell}
\end{eqnarray}
by using the relations
$\Gamma_{i\uparrow j\uparrow}^{i\uparrow j\uparrow}=\Gamma_{i\downarrow j\downarrow}^{i \downarrow j\downarrow}$ and
$\Gamma_{i\uparrow j\downarrow}^{i\uparrow j\downarrow}=\Gamma_{i\downarrow  j\uparrow}^{i\downarrow j\uparrow}$,
 see the Supporting Information (SI) and Ref.~\cite{Kutzelnigg2010} for details.
 

\section{results}

Calculations are performed on three square lattices with tilted periodic boundary conditions,
namely the 18-site lattice with lattice vectors $(3,3), (3,-3)$, the 20-site lattice with lattice vectors $(4,-2), (-2,-4)$, and the 24-site lattice with lattice vectors $(5,1), (-1,-5)$. The supercell of the 24-site lattice is not square, but the underlying lattice still is. This 24-site lattice has been discussed comprehensively by Betts~\cite{betts1999} with regard to its finite-size properties, which are indeed favourable, having a low topological imperfection as defined by Betts.
The above three tilted lattices are presented in~\cite{yun2021} (see Fig.1 therein).

\subsection{The two-hole system}
\begin{figure}[h!]
\includegraphics[width=0.35\textwidth]{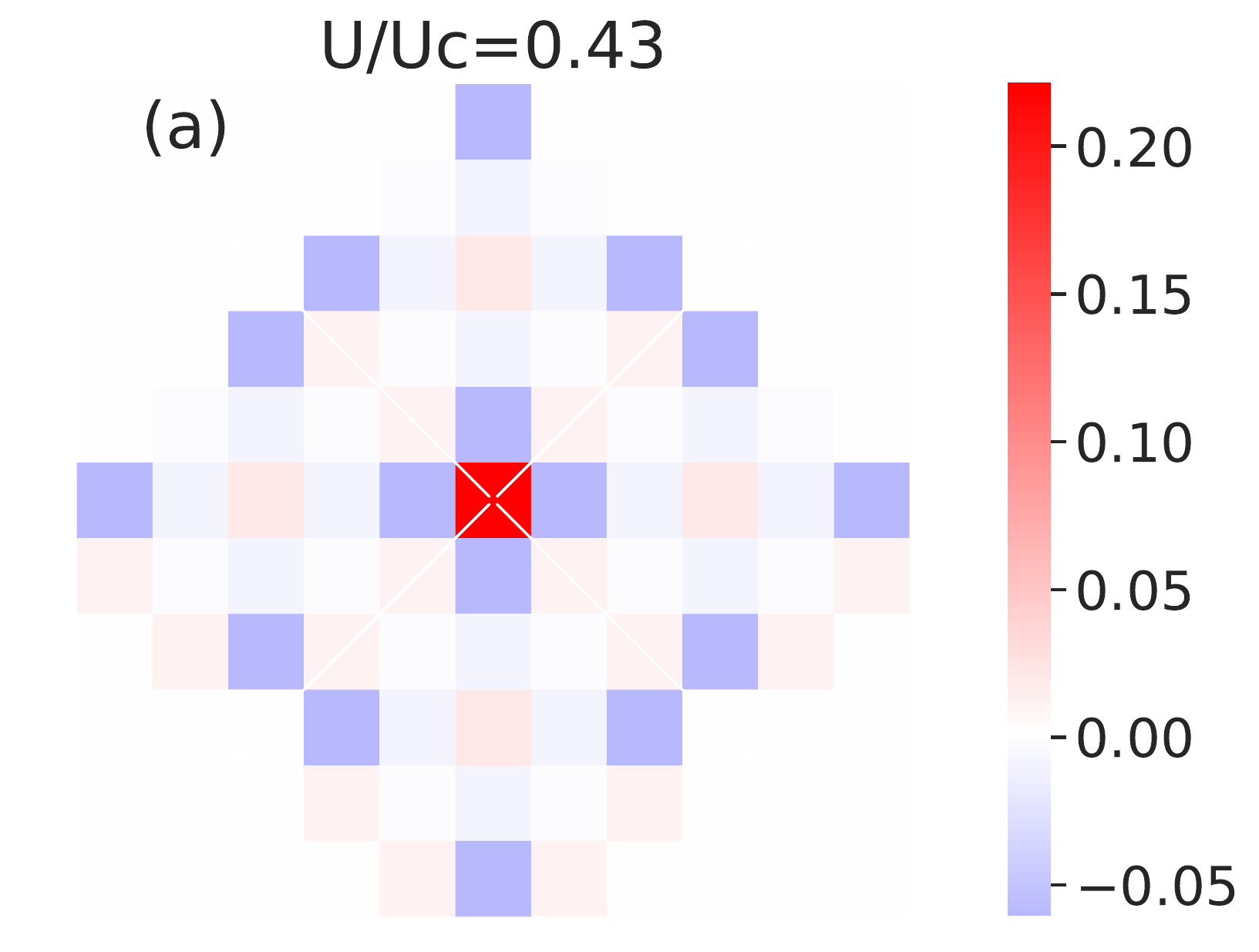}
\includegraphics[width=0.35\textwidth]{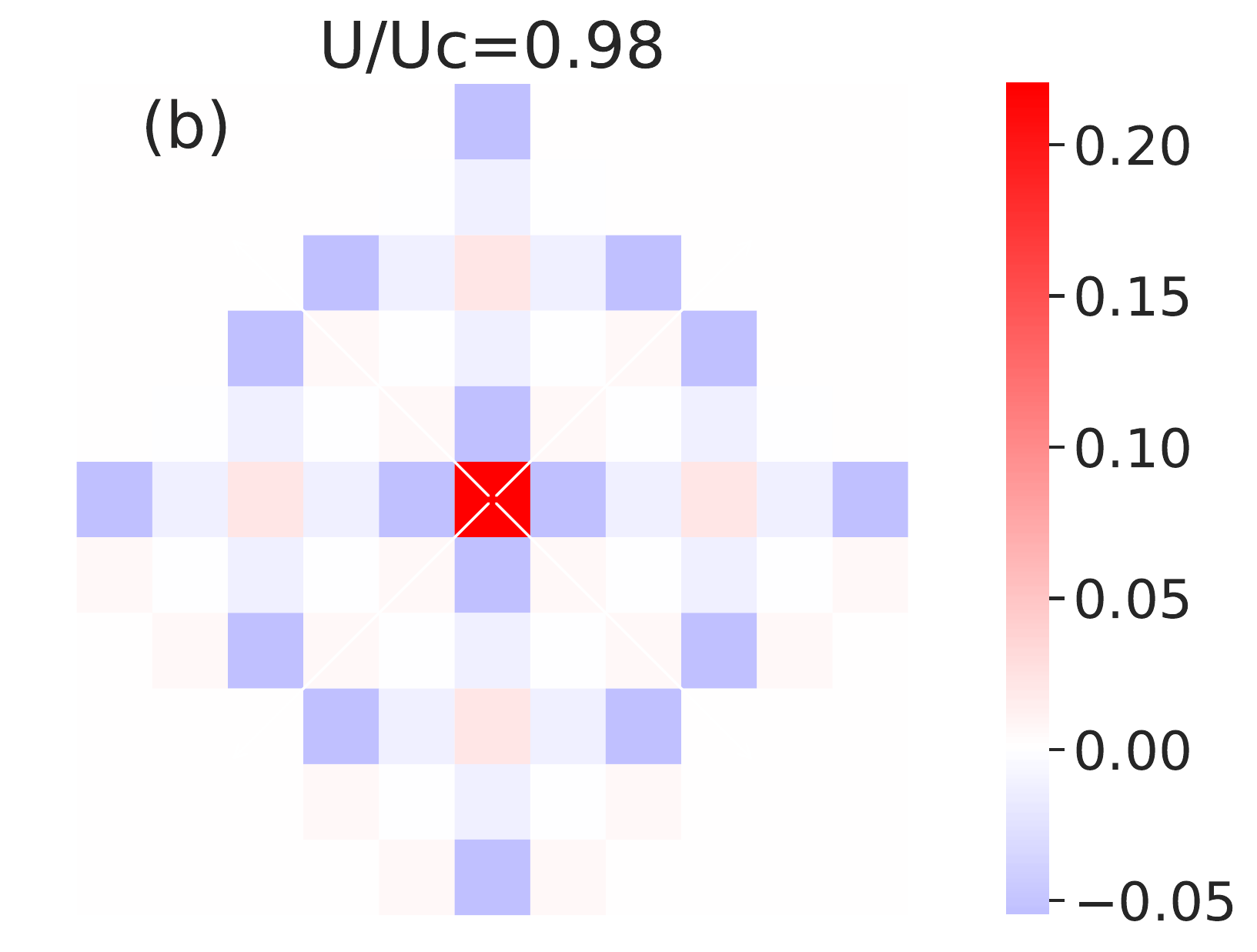}
\includegraphics[width=0.35\textwidth]{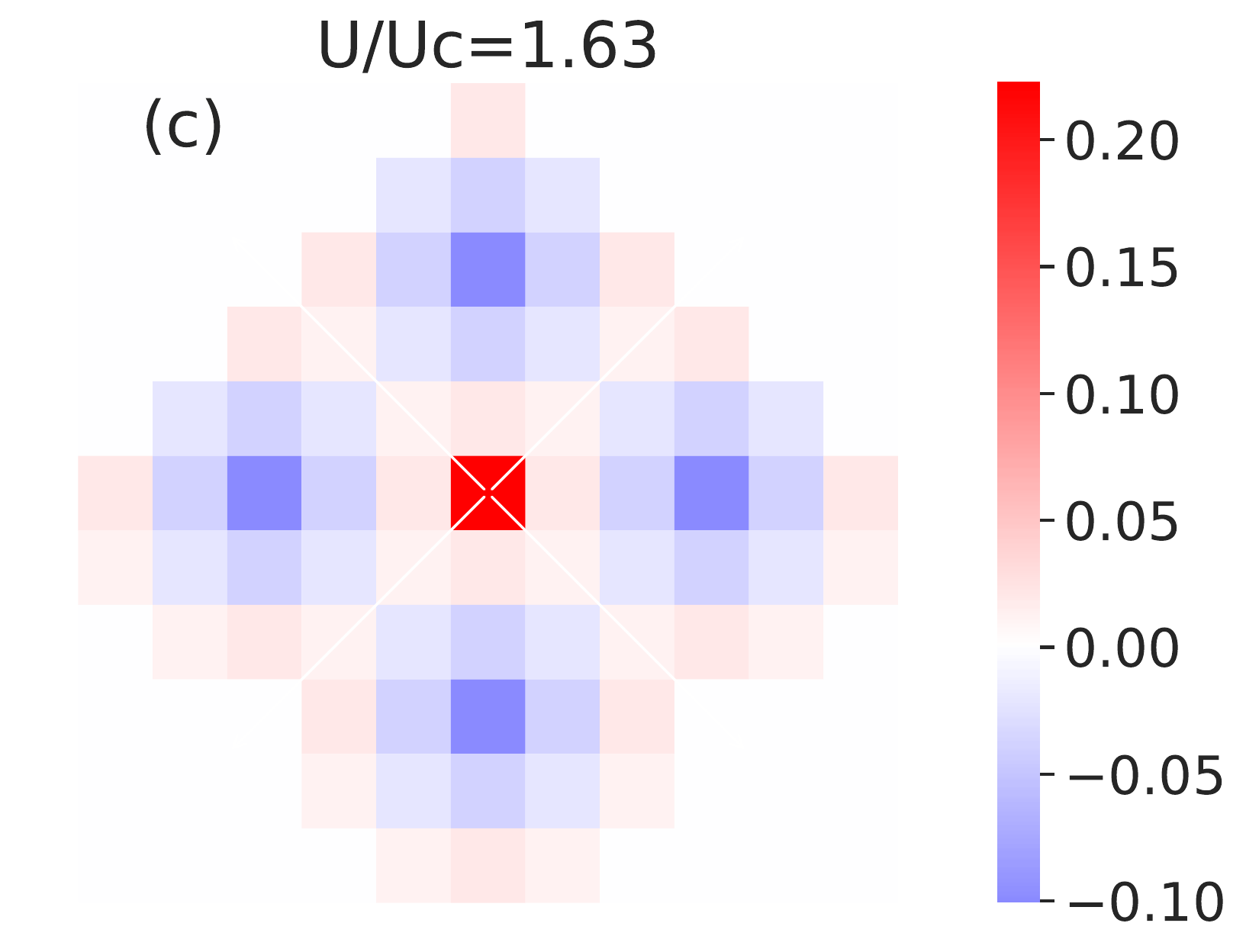}
\includegraphics[width=0.35\textwidth]{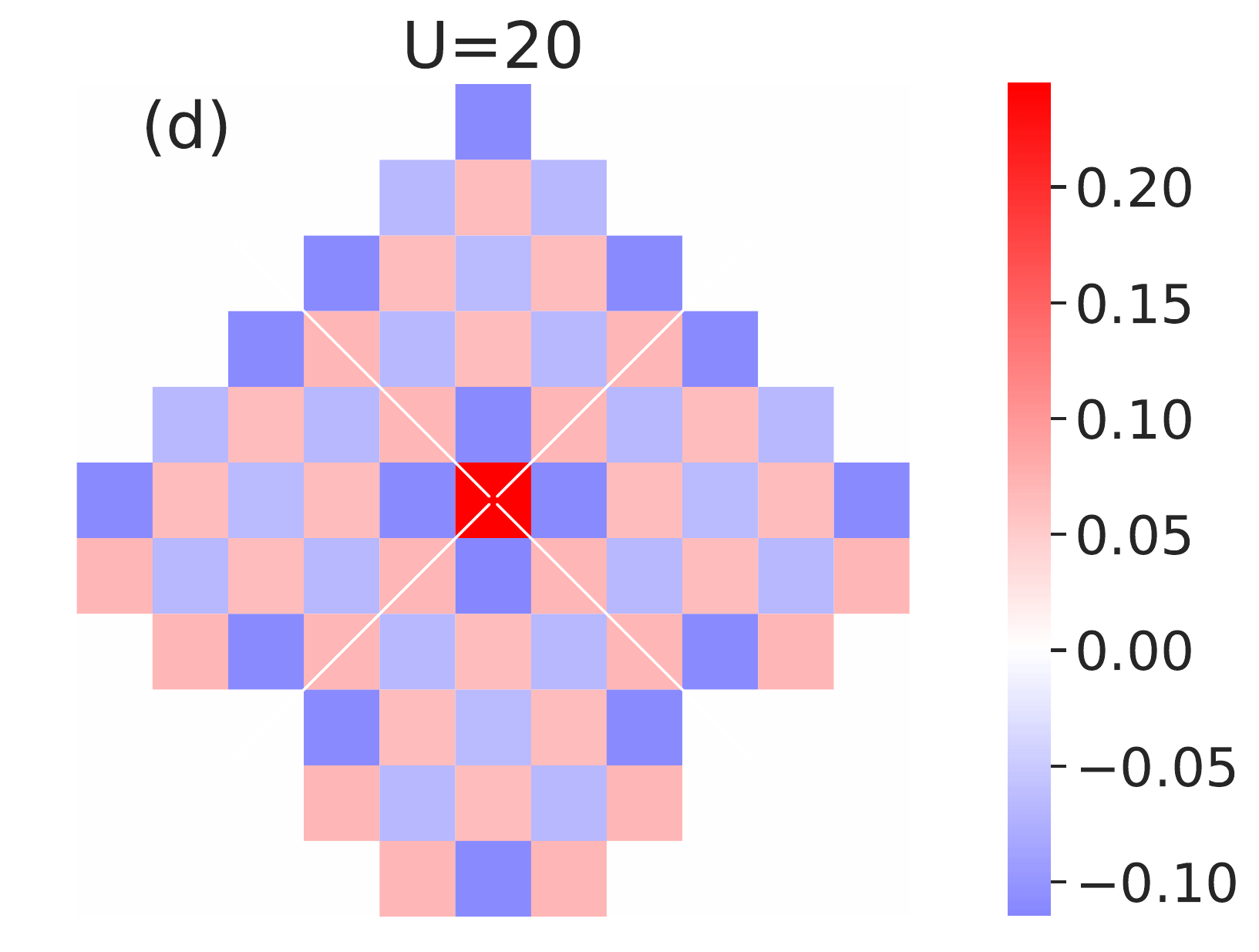}
\begin{flushleft}
\caption{\label{s18withtwoholes} Spatial spin distributions given by $\langle \hat{S}_z(i)\hat{S}_z(j)\rangle$ on the 18 sites lattice  for  two-hole (a-c) and half-filled (d) systems. The critical interaction strength is $U_c=92$ and the site $i$ is set in the center.}
\end{flushleft}
\end{figure}

\begin{figure}[h!]
\centering
\includegraphics[width=0.35\textwidth]{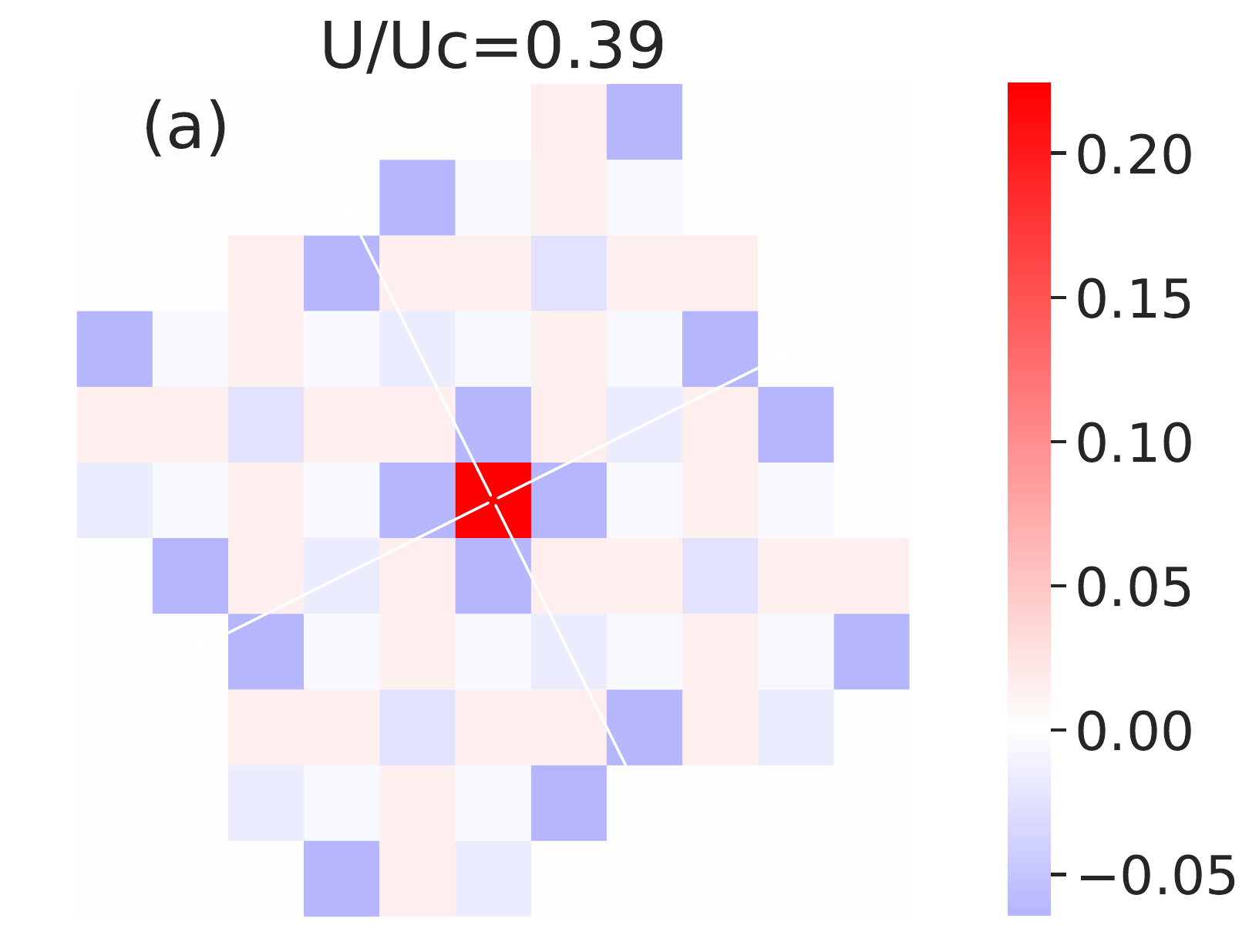}
\includegraphics[width=0.35\textwidth]{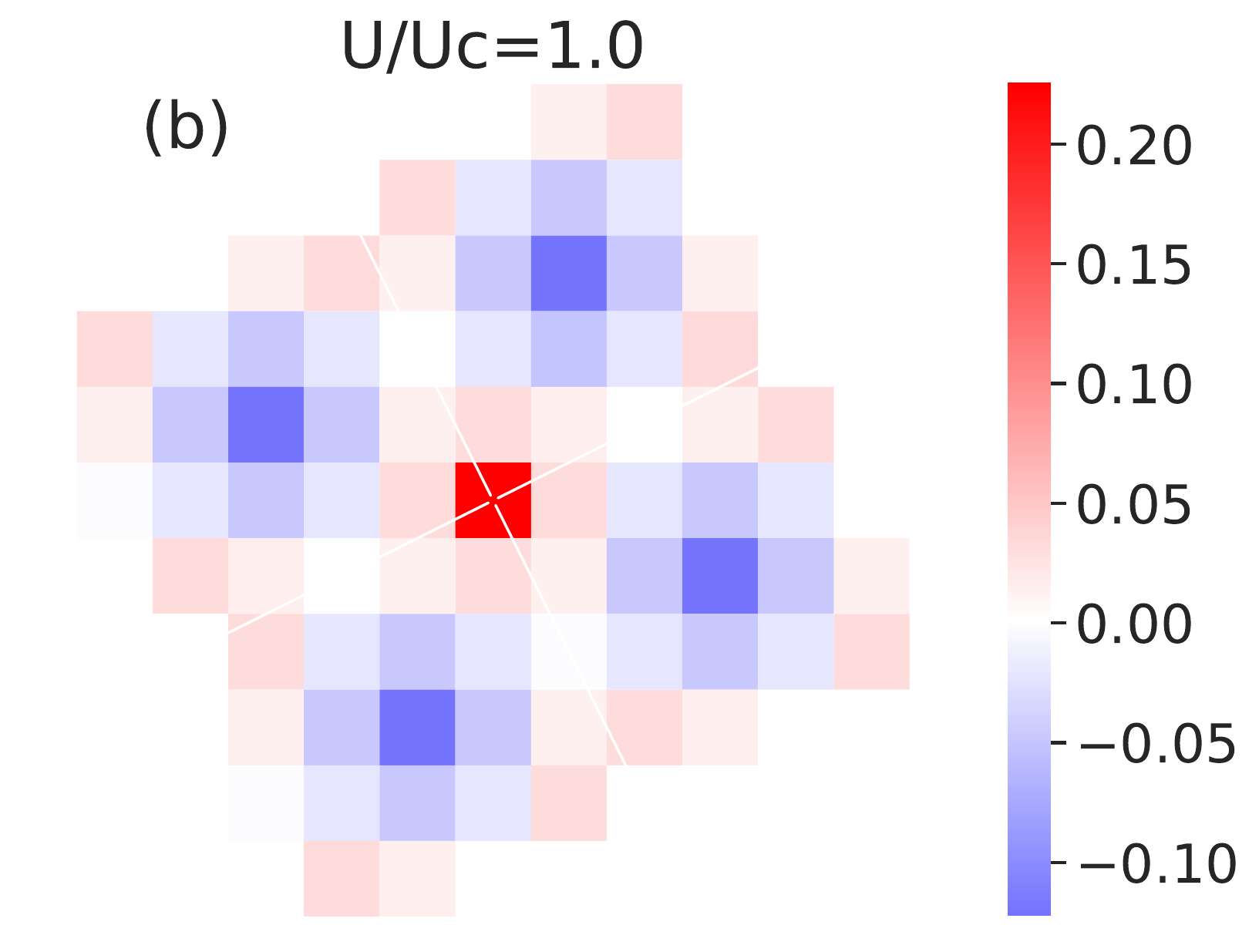}
\includegraphics[width=0.35\textwidth]{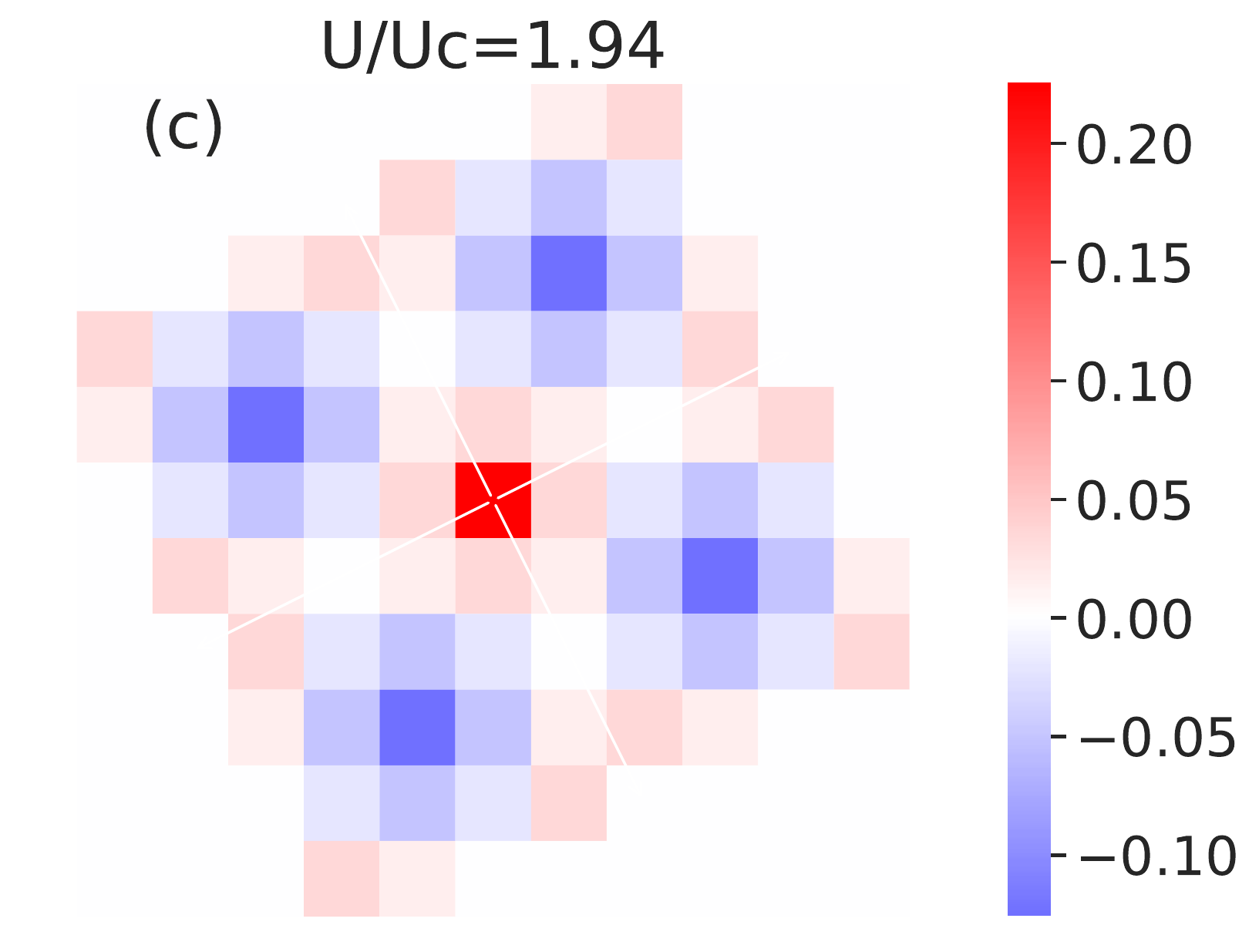}
\includegraphics[width=0.35\textwidth]{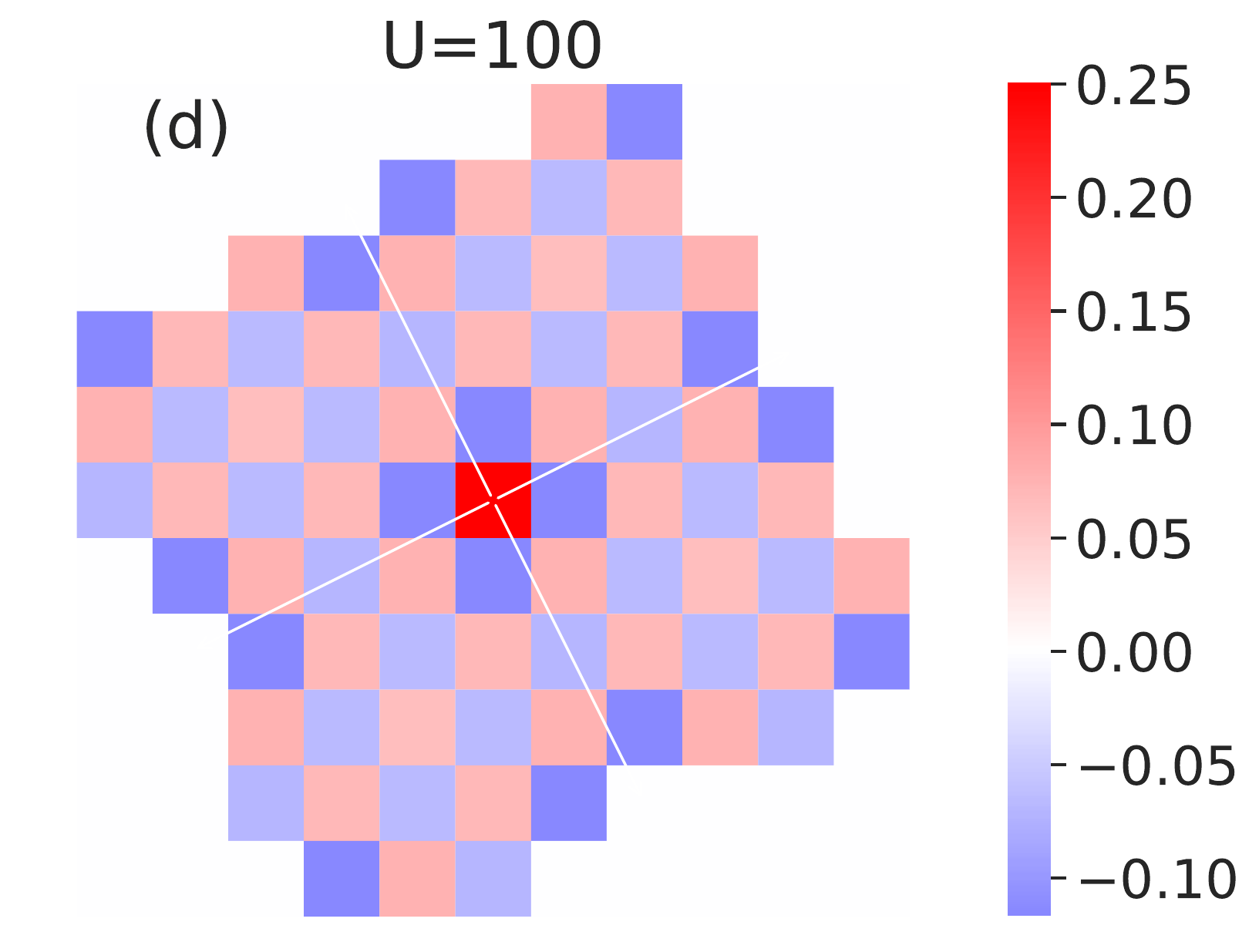}
\begin{flushleft}
\caption{\label{s20withtwoholes}Spatial spin distributions given by $\langle \hat{S}_z(i)\hat{S}_z(j)\rangle$ on the 20 sites lattice for two-hole (a-c) and half-filled (d) systems.The critical interaction strength is $U_c=103$.}
\end{flushleft}
\end{figure}

\begin{figure}[h!]
\centering
\includegraphics[width=0.35\textwidth]{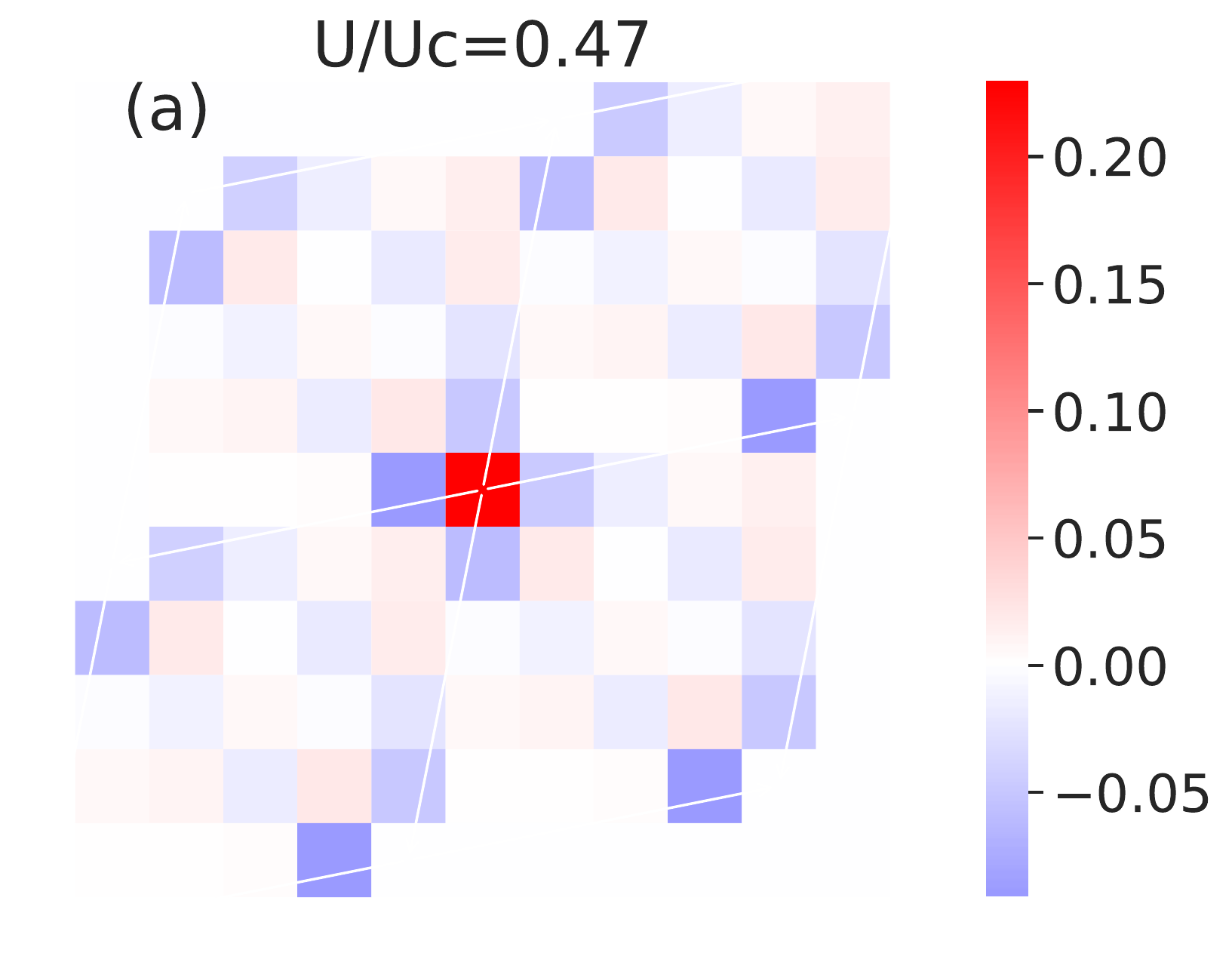}
\includegraphics[width=0.35\textwidth]{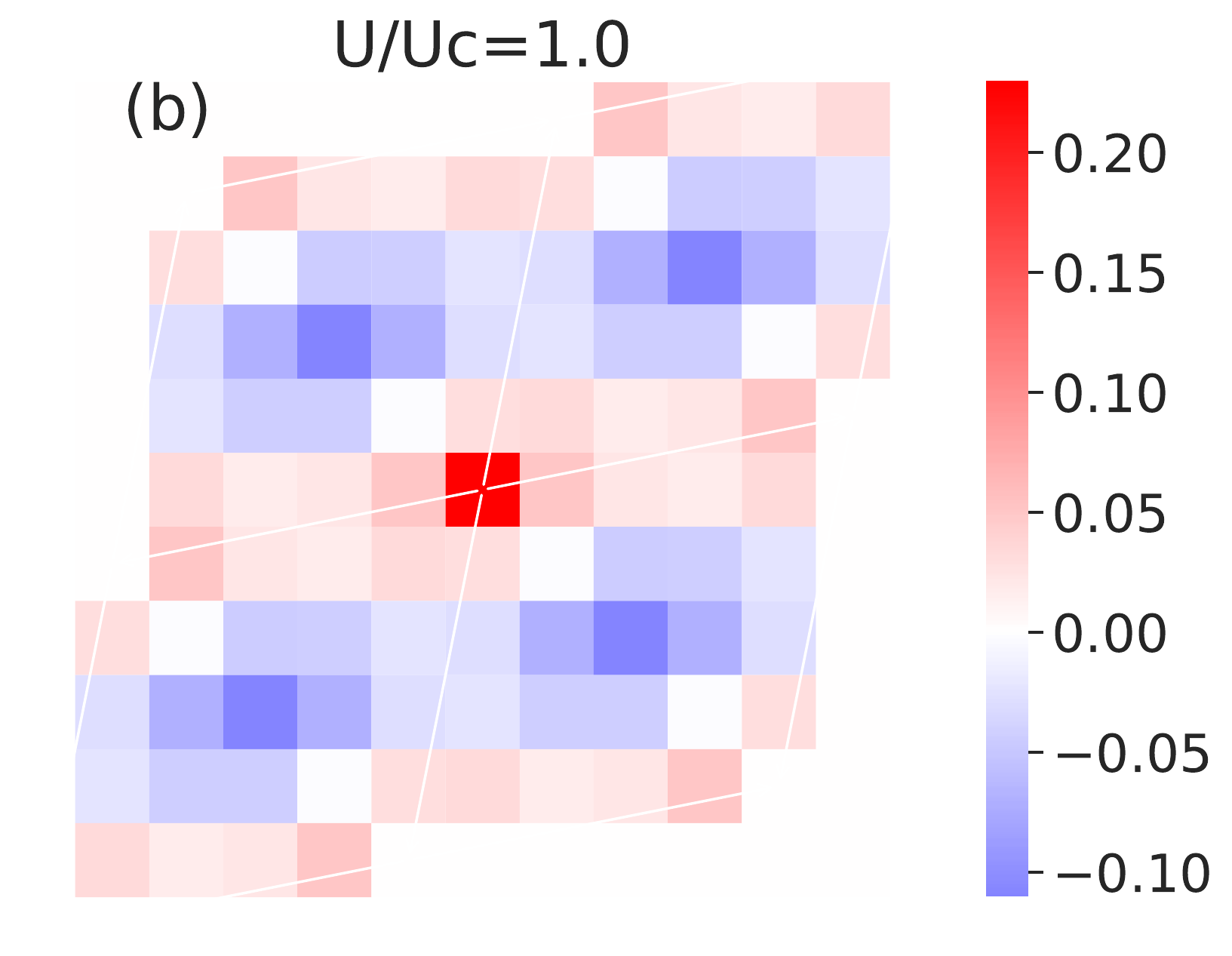}
\includegraphics[width=0.35\textwidth]{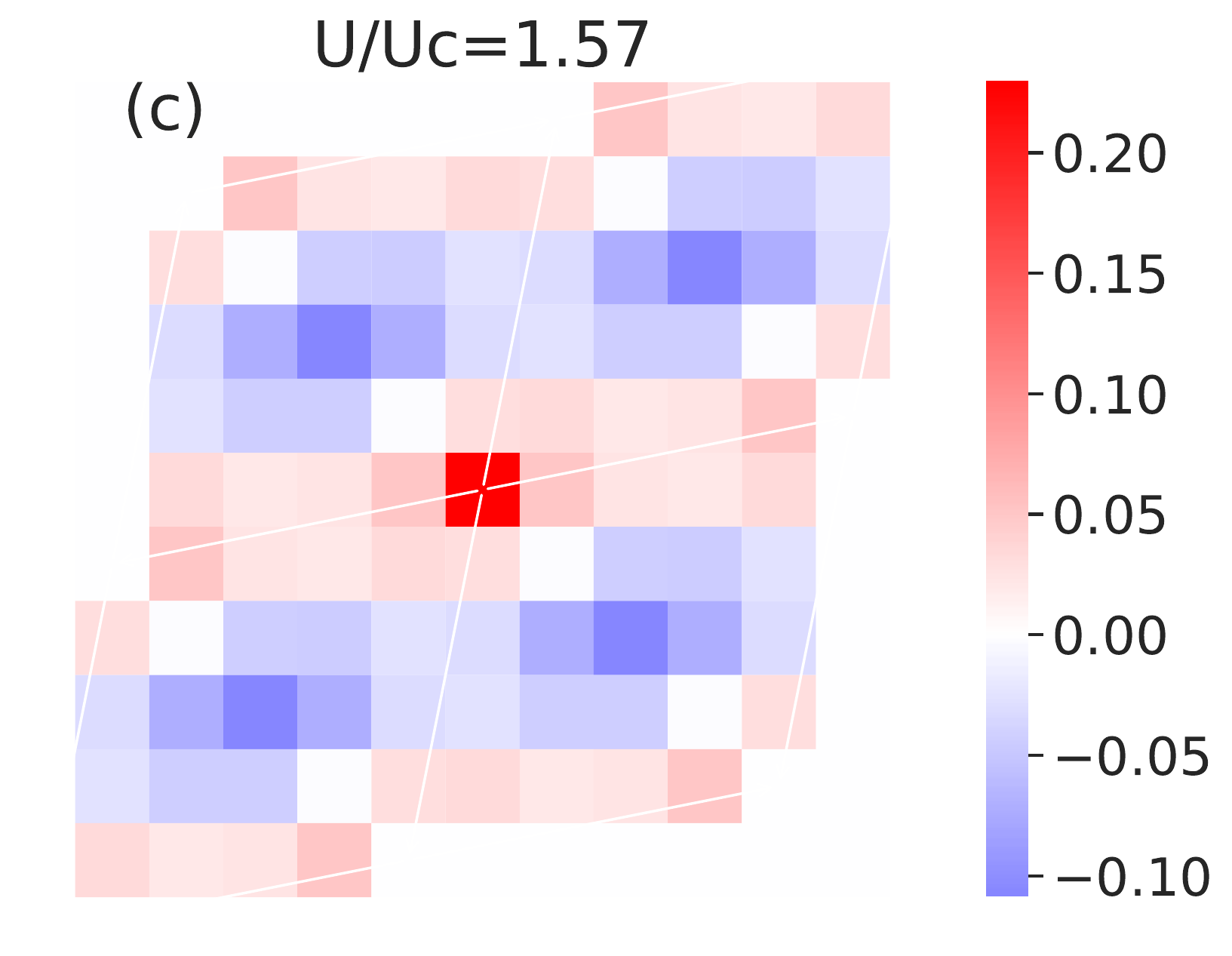}
\includegraphics[width=0.35\textwidth]{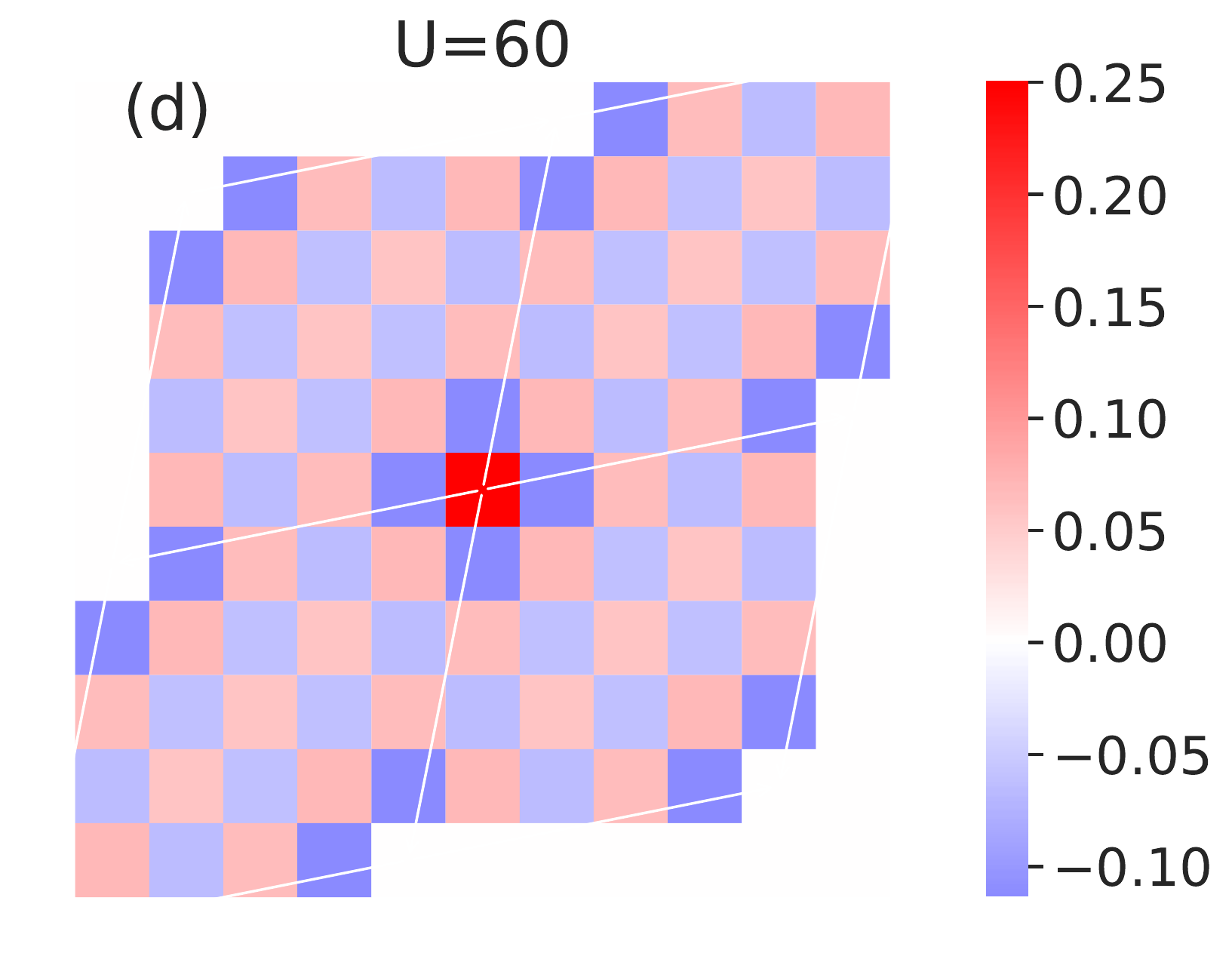}
\begin{flushleft}
\caption{\label{s24withtwoholes} Spatial spin distributions given by $\langle \hat{S}_z(i)\hat{S}_z(j)\rangle$ on the 24 sites lattice for two-hole (a-c) and half-filled (d) systems. 
The critical interaction strength is $U_c=127$.}
\end{flushleft}
\end{figure}

 \begin{figure}[h!]
\centering
\includegraphics[width=0.32\textwidth]{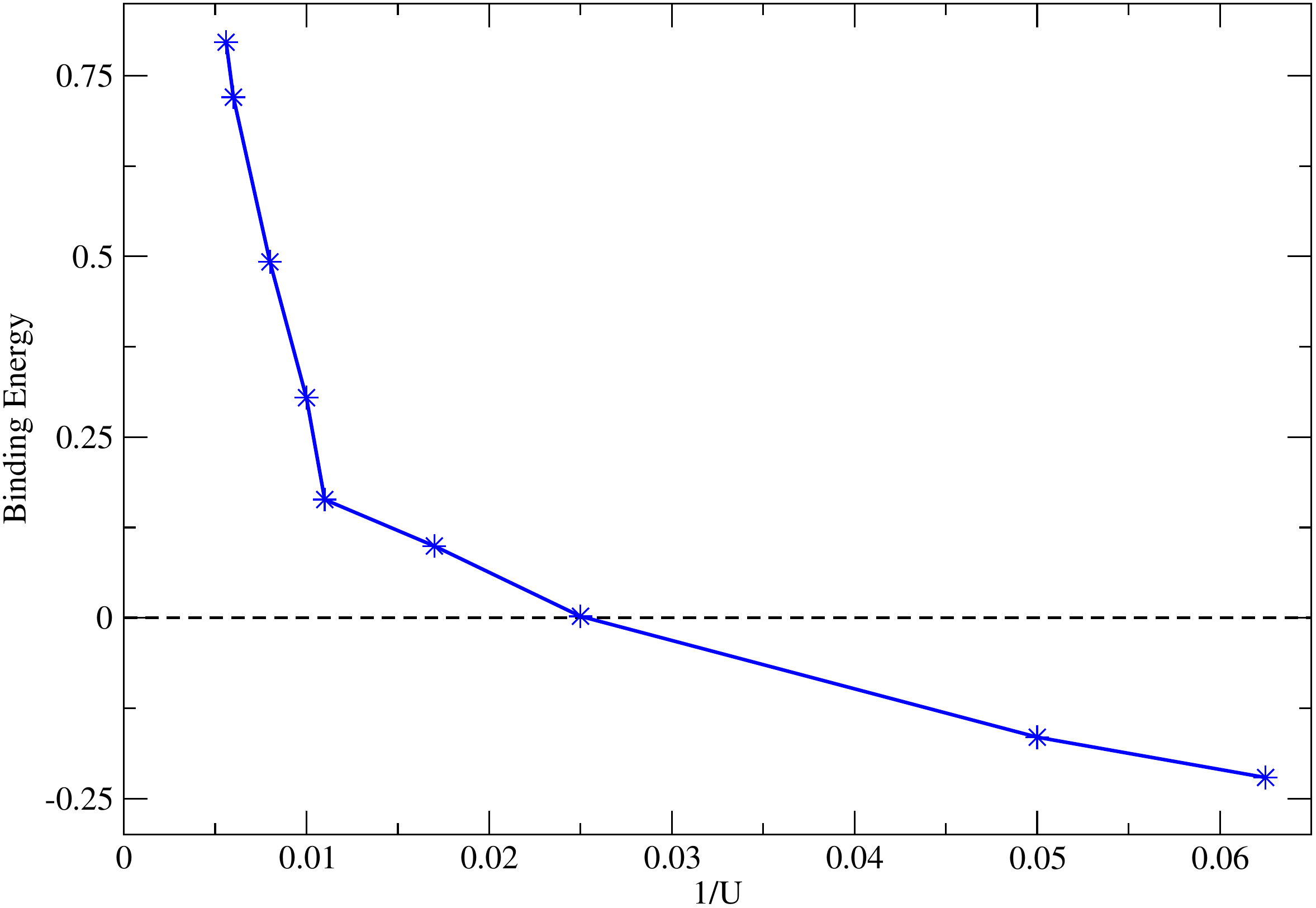}
\begin{flushleft}
\caption{\label{binding energy} Binding energy of two holes vs. 1/U on 18 sites}
\end{flushleft}
\end{figure}

 \begin{figure}[h!]
\centering
\includegraphics[width=0.35\textwidth]{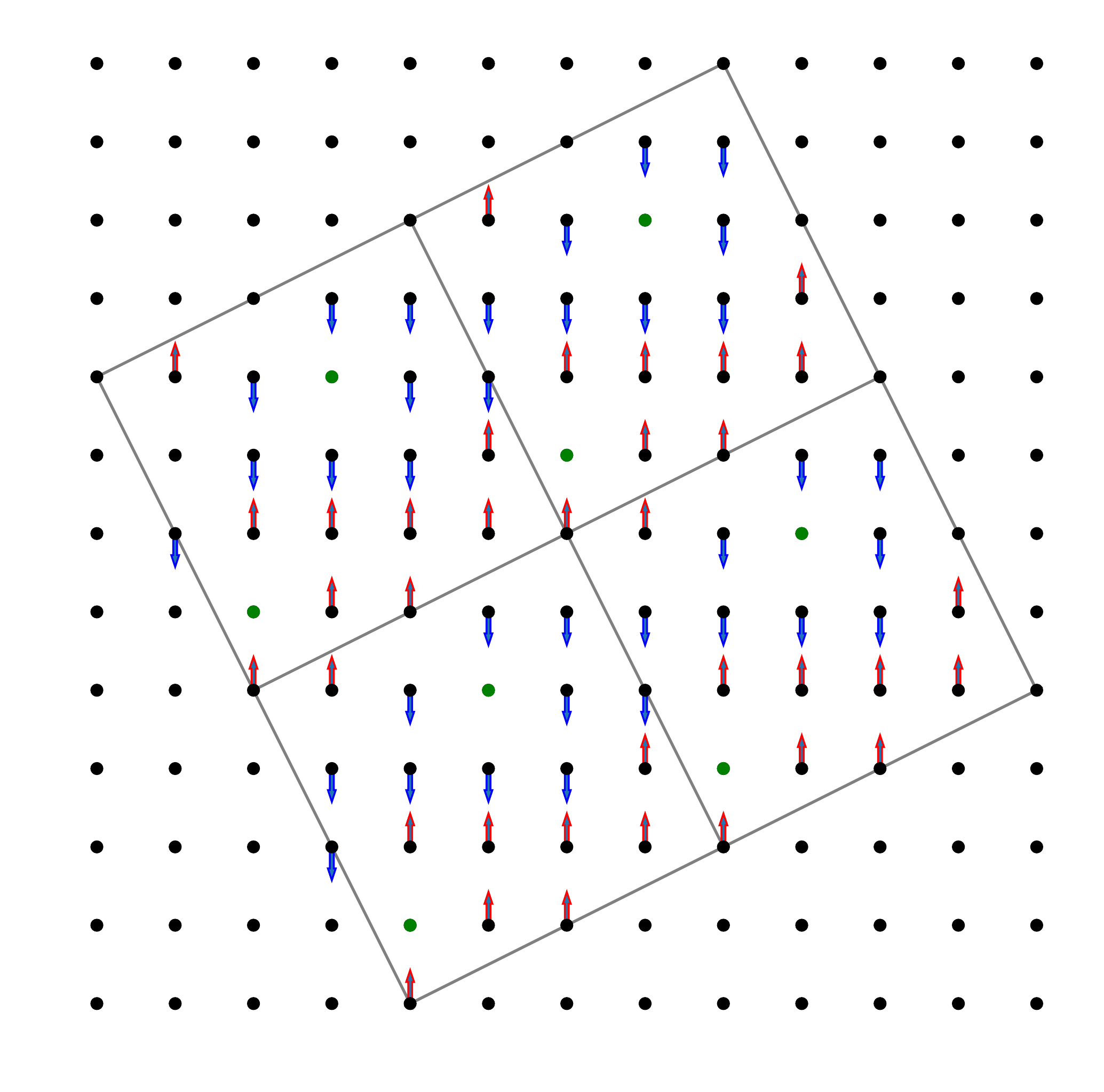}
\begin{flushleft}
\caption{\label{det} Spin distributions of the Slater determinant with the  most weight in the ground state of the 20 sites lattice with two holes. The red (blue) arrows represent spin up (down), and the green dots represent a hole. The lattice cutout is composed of four supercells.}
\end{flushleft}
\end{figure}

\begin{figure}[h!]
\centering
\includegraphics[width=0.35\textwidth]{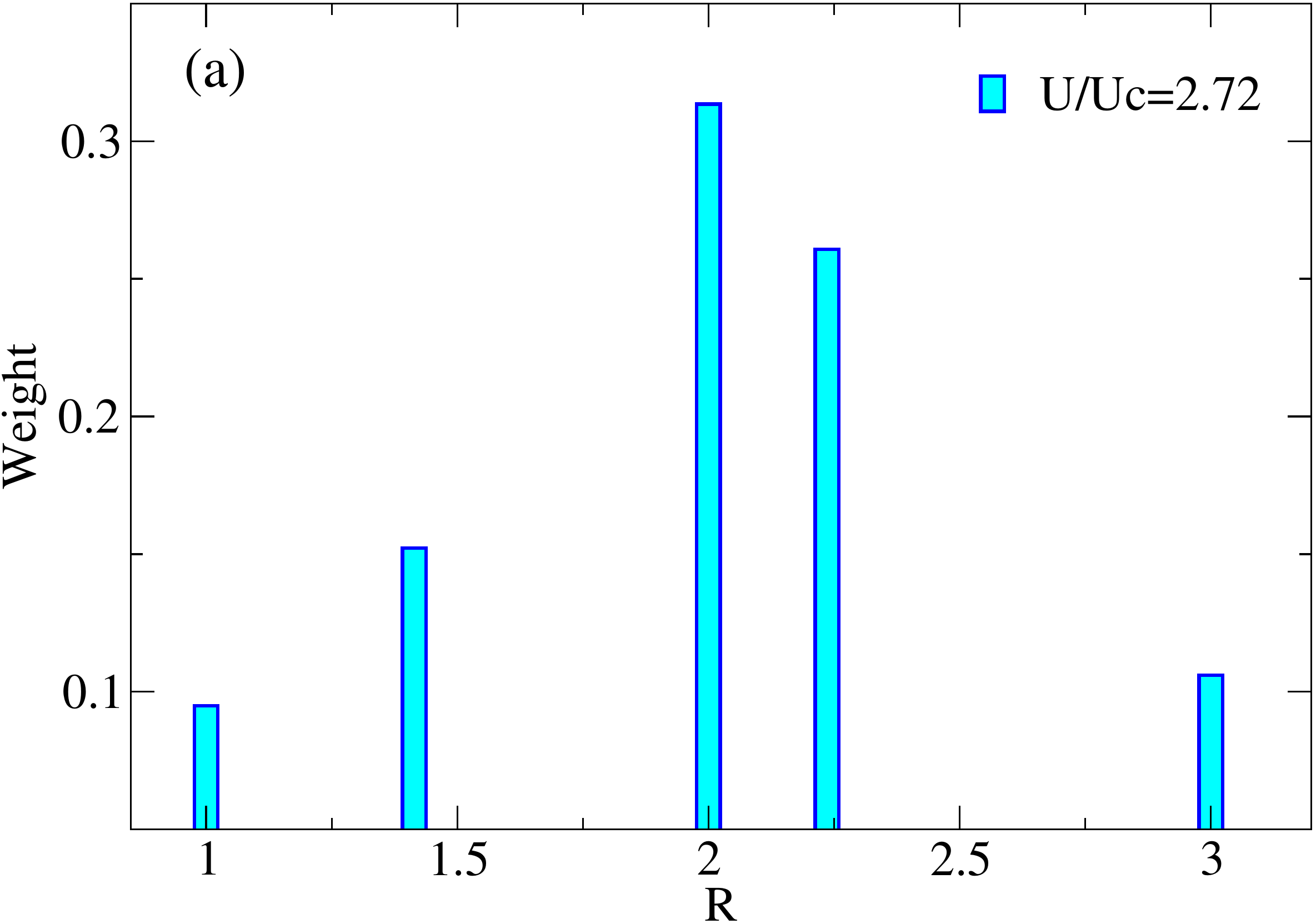}
\includegraphics[width=0.35\textwidth]{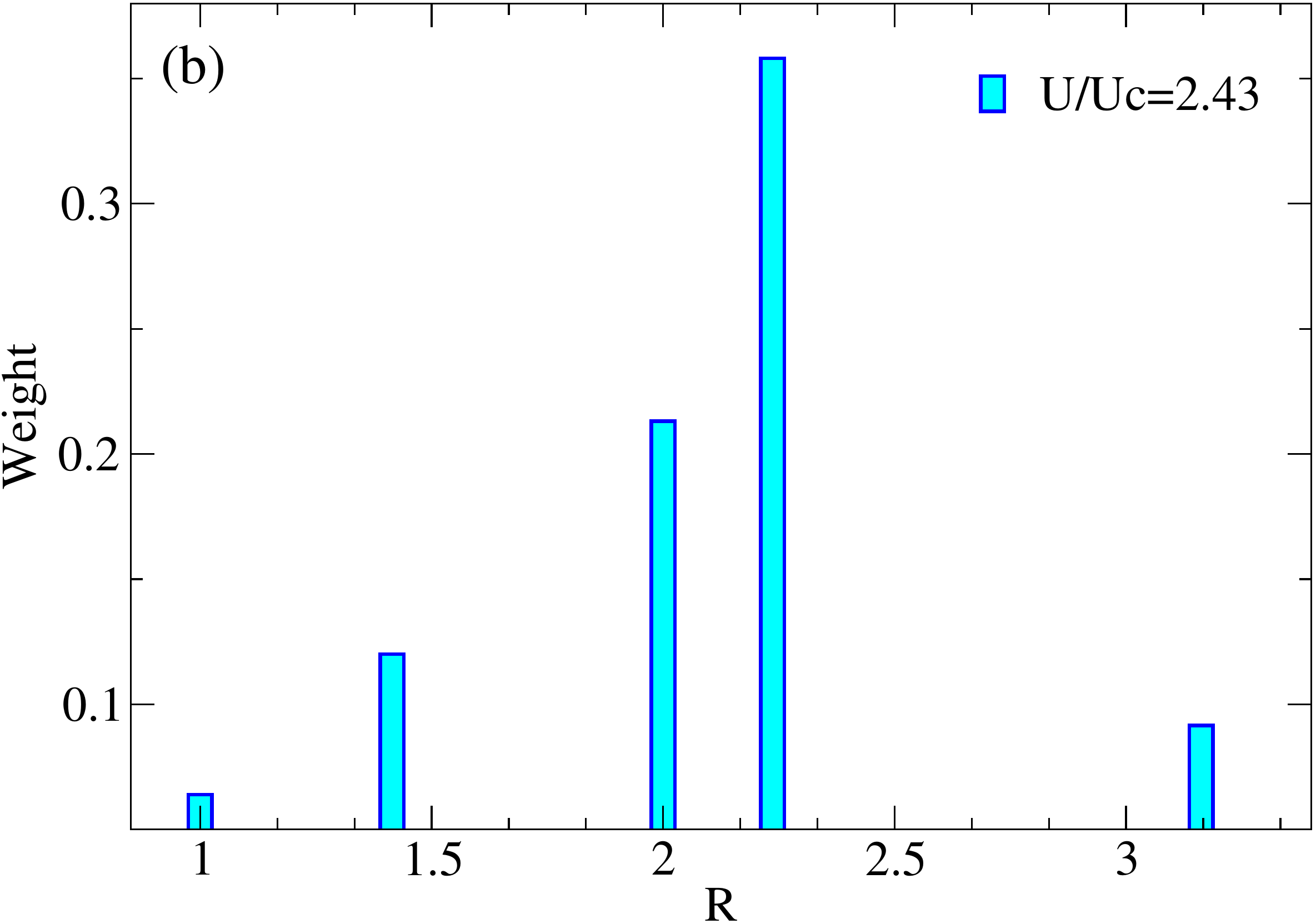}
\includegraphics[width=0.35\textwidth]{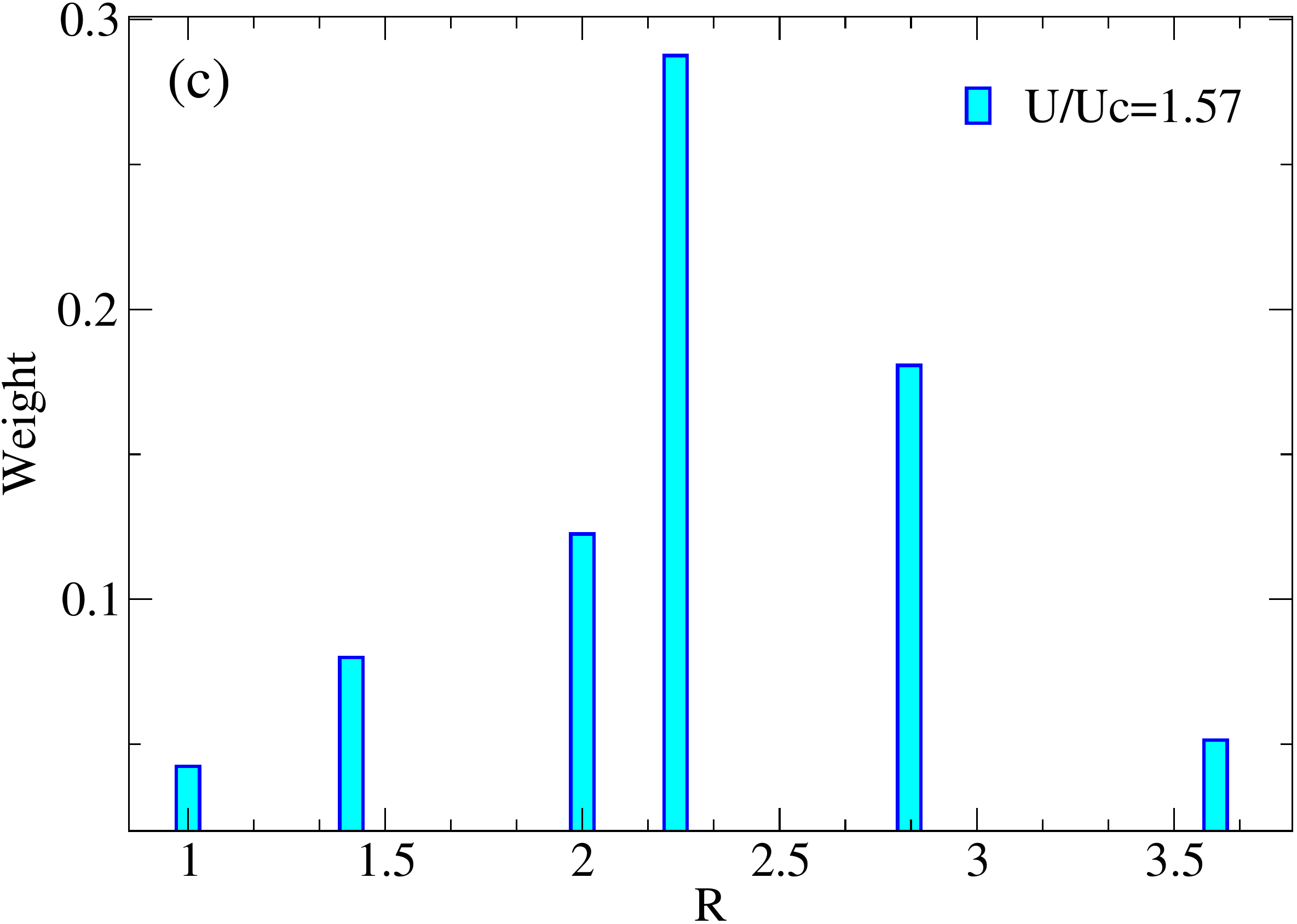}
\begin{flushleft}
\caption{\label{distance} Statistical distribution of the distance between two holes on 18-(a), 20-(b) and 24-site(c) lattices. The number of considered SDs in the ground-state wave function considered for the analysis are $5 \times 10^{5}$ (0.93\% of total occupied SDs in the FCIQMC ground state estimate),
$8 \times 10^{5}$(0.85\% of occupied SDs), and $10^{7}$(0.76\% of occupied SDs), respectively. Please note that the ground-state wave function is not normalized and the weight shown in  the figure is a relative value.}
\end{flushleft}
\end{figure}

The two-hole system has been investigated in our previous work~\cite{yun2021}, where we find the total spin of the ground state is always zero for all the different lattices. This supports the early conclusions based on
exact diagonalization~\cite{riera1989} and analytical studies of arbitrarily large systems~\cite{wen1989}.

\begin{table*}[ht]
  \centering \caption{\label{tab1} The ground state $S=0$ spin-spin correlations $\langle {\bf S}(i)\cdot {\bf S}(j)\rangle$ obtained from FCIQMC calculations on the 18-site square lattice Hubbard model with 16 electrons for different values of $U$. For comparison the values from the exact diagonalization(ED) calculation taken from Ref.~\cite{becca2000} are also shown. 
  $i$ and $j$ are the lattice site indices, and $R$ is their distance. }
\vspace{0.2cm}
 \begin{tabular} {|c|cc|cc|cc|cc|cc|}  \hline
 \diagbox[width=2.50cm,height=1.50cm,linewidth=1pt] {U}{$\langle {\bf S}(i)\cdot {\bf S}(j)\rangle$}{R}& \multicolumn{2}{c}{1} \vline  & \multicolumn{2}{c}{$\sqrt{2}$} \vline & \multicolumn{2}{c}{2} \vline & \multicolumn{2}{c}{$\sqrt{5}$}\vline & \multicolumn{2}{c}{3} \vline \\ [0.3cm] 
 \cline{1-11}
   & 2-RDM & ED & 2-RDM & ED  & 2-RDM & ED & 2-RDM & ED & 2-RDM & ED \\[0.2cm] 
  20  & $-0.1910(1)$ &  $-0.19096$  & $0.0448(1)$ &  $0.04473$ &  $-0.0154(1)$ & $-0.01541$ & $-0.01401(1)$ & $-0.01398$ & $0.0493(1)$ & $0.04931$   \\ [0.2cm] 
  40   & $-0.1820(1)$ & $-0.18197$   & $0.0341(1)$ & $0.03417$ &  $-0.0274(1)$ &    $-0.02739$ &  $-0.0062(1)$ &   $-0.0061$  &  $0.0619(1)$ &  $0.06188$   \\ [0.2cm]
  \hline
  \end{tabular}
  \begin{flushleft}
  \end{flushleft}
 \end{table*}


To check the performance of our algorithm, we first perform calculations on the 18-site lattice, where Lanzcos-based ED results are available~\cite{becca2000}. In Table~\ref{tab1}, the results of the spin-spin correlation, $\langle \hat{\bf S}(i)\cdot \hat{\bf S}(j)\rangle$, on  the 18-site lattice for $U=20$ and $40$ are presented in comparison with the ED results.  In our simulation we use a  time step of $\tau=0.001$,  $N_{w}=5\cdot10^7$  number of  walkers and treat the $5\cdot 10^4$ most populated states deterministically. 
As shown in Table~\ref{tab1}, the two results  agree very well for different $R$ (the distance between two different sites $i$ and $j$). 

In order to get a visual impression of the spatial spin distribution, we plot the pair correlation function $\langle \hat{S}_z(i)\hat{S}_z(j)\rangle$ for different interaction strengths and number of holes
in a lattice cutout composed of four supercells, as being presented in Figures~\ref{s18withtwoholes},~\ref{s20withtwoholes} and~\ref{s24withtwoholes}, respectively for the 18, 20 and 24-sites lattices. The lattice site $i$ is set in the center of the cutout.

On the 18-site lattice, the spatial spin distribution in the half-filling case shows a clear anti-ferromagnetic pattern as shown in Fig.~\ref{s18withtwoholes}(d). 
For two-hole systems, we still find some anti-ferromagnetic features when $U$ is not too large, as the cases showing in  Fig.~\ref{s18withtwoholes}(a) and \ref{s18withtwoholes}(b), where spins on neighboring sites are still anti-parallel. However, when $U$ is larger than a certain critical value ($U_c$), as being presented in Fig.~\ref{s18withtwoholes}(c),  the neighboring spins become parallel, and we find some kind of ferromagnetic domains structure. 
It is interesting to find that the critical $U_c$  is very close to that of Nagaoka ferromagnetism in one-hole systems. We therefore do not try to determine these critical $U_c$ values, but rather use those $U_c$'s determined for the Nagaoka ferromagnetisation as references. Such transitions of the spatial spin distribution patterns are also found on the 20- and 24-site lattices, see Fig.~\ref{s20withtwoholes} and Fig.~\ref{s24withtwoholes},  where we also find ferromagnetic domains when $U$ is large enough.   
When the lattice size increases, the size of the ferromagnetic domains becomes larger, and their shape changes.

 The fact that critical $U_c$'s are close to those of Nagaoka ferromagnetism in one-hole systems, provides a physical picture that holes tend to separate from each other. 
 This picture,  originally suggested by Tian~\cite{tian1991}, is now tested and verified based on the analysis of Slater determinants of the ground-state.  For the two-hole systems, we have also found that for large $U$'s the binding energies are positive as shown in Fig.~\ref{binding energy}.
 The binding energy of two holes is defined as $\delta=(E_{2}-E_{0})-2(E_{1} - E_{0})=E_{2}-2E_{1}+E_{0}$, where $E_{n}$ is the ground-state energy of the $n$-hole system. 
 The positive binding energy implies that two holes tend to separate, rather than to bind for large $U$'s. In this case, it is reasonable to assume that each hole carries a halo of ferromagnetic texture. 

To further confirm the picture of the hole separations we have looked at the  spin distribution of the most populated SD in the ground-state of the 20-site lattice. 
For $U=200$, the spin distribution of the SD with highest weight is plotted in Fig.~\ref{det}, where we find that the distance between two holes is the farthest, $\sqrt{10}$, and spins around each hole are the same. The analysis of the $8\times 10^5$ most populated SDs shows that the most common hole-distance is not $\sqrt{10}$, but $\sqrt{5}$ instead (as shown in Fig.~\ref{distance}(b)). Similarly on the 24-site lattice, $\sqrt{5}$ is the most common hole-distance (Fig.~\ref{distance}(c)), and on the 18-site lattice, the distance $2$ is slightly more common than $\sqrt{5}$ (Fig.~\ref{distance}(a)). Ferromagnetic domains appear on finite lattices when holes are far away from each other. For systems with a fixed number of holes, 
the possibility of the formation of ferromagnetic domains increases with the system size, which means that the ferromagnetic domain structure favors a low hole density.

\subsection{The three-hole system}

\begin{figure}[h!]
\centering
\includegraphics[width=0.35\textwidth]{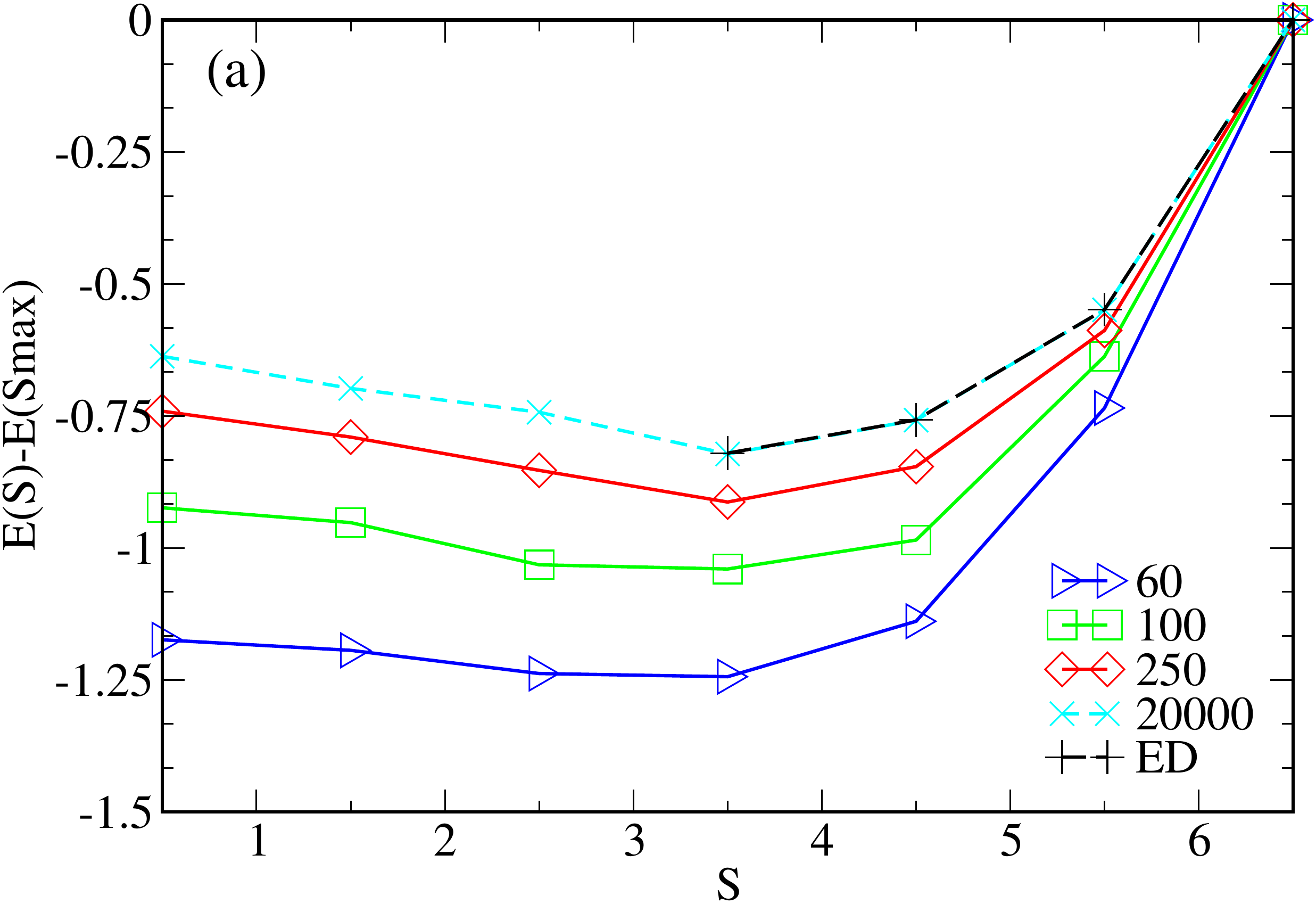}
\includegraphics[width=0.35\textwidth]{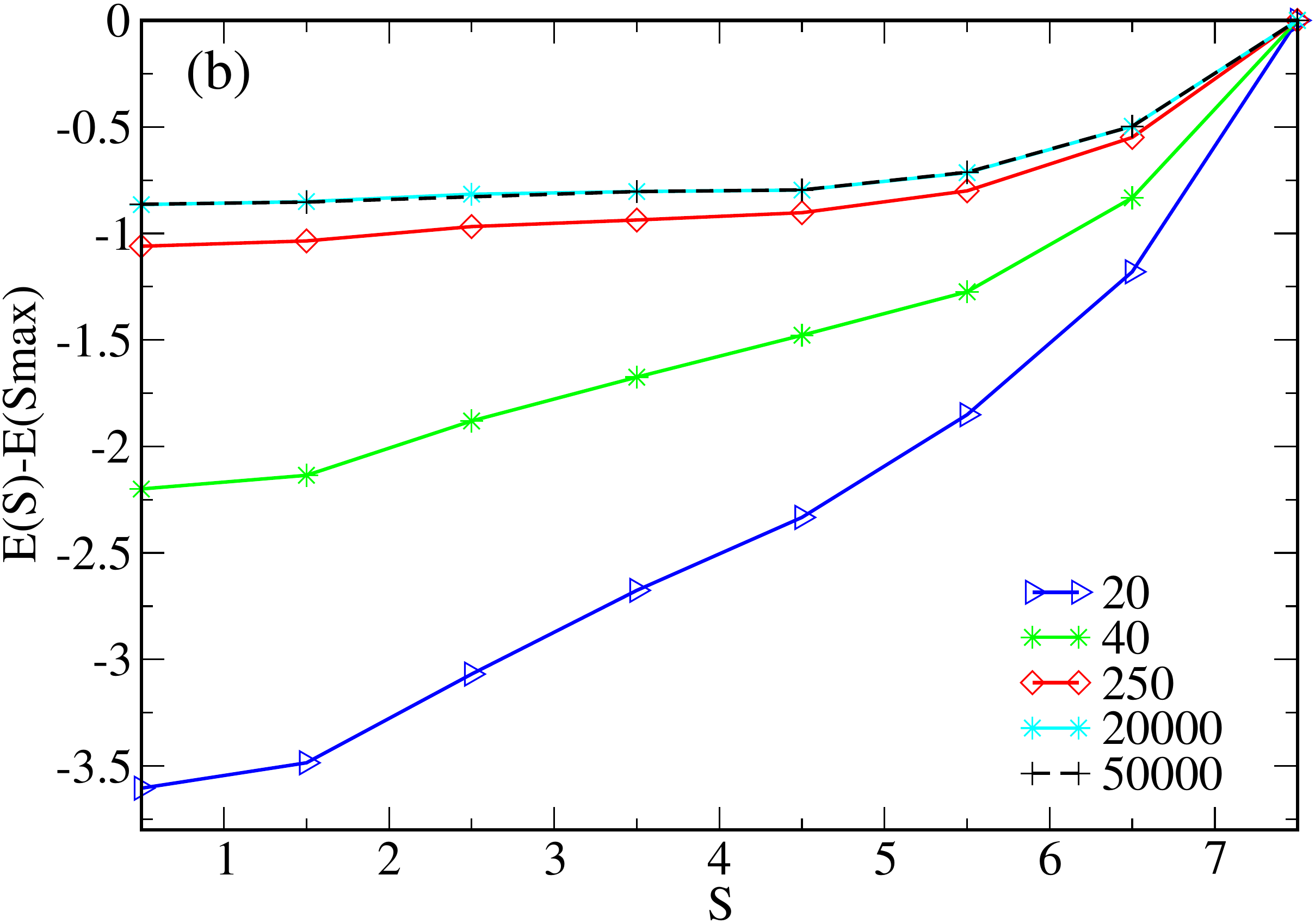}
\includegraphics[width=0.35\textwidth]{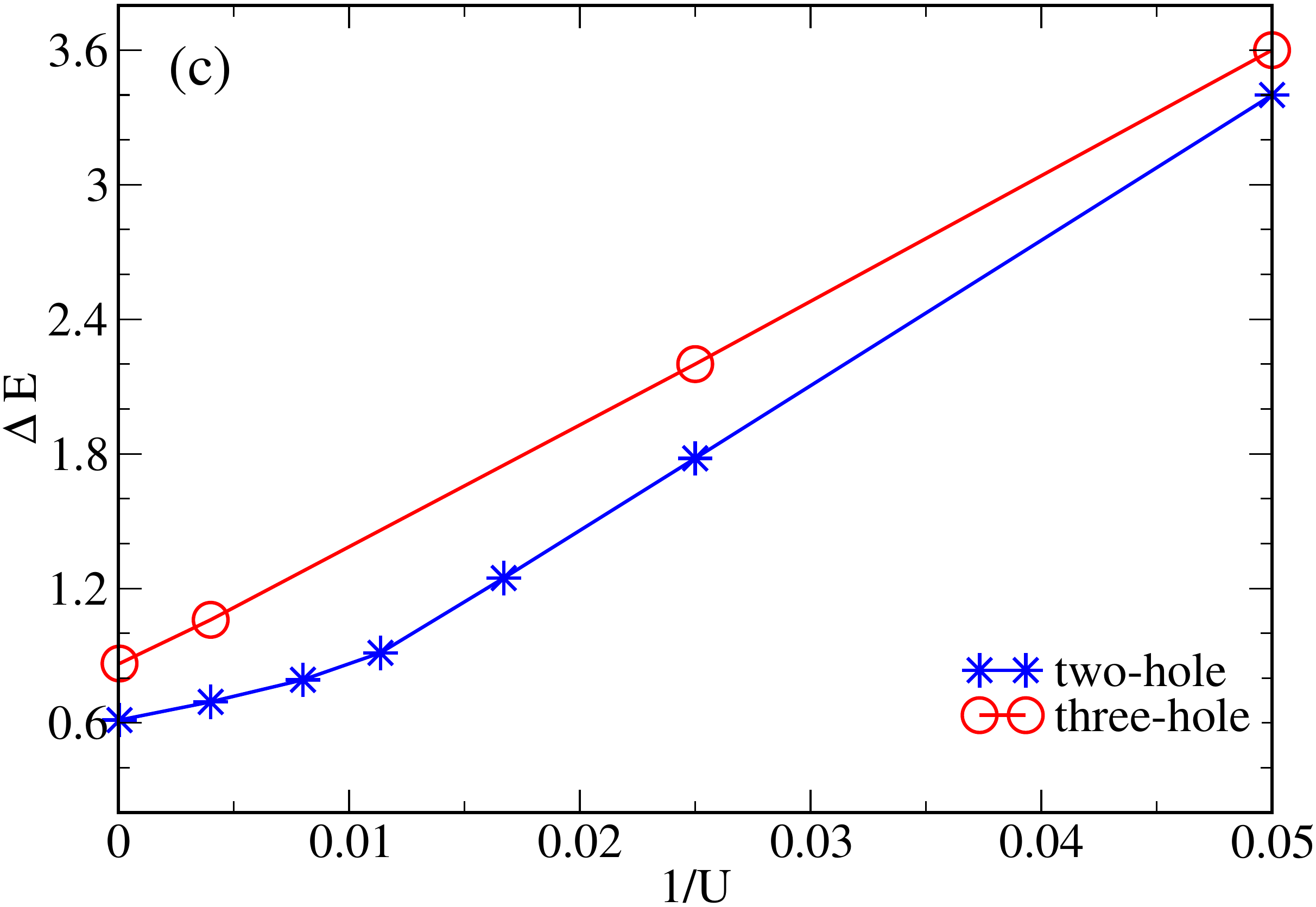}
\includegraphics[width=0.35\textwidth]{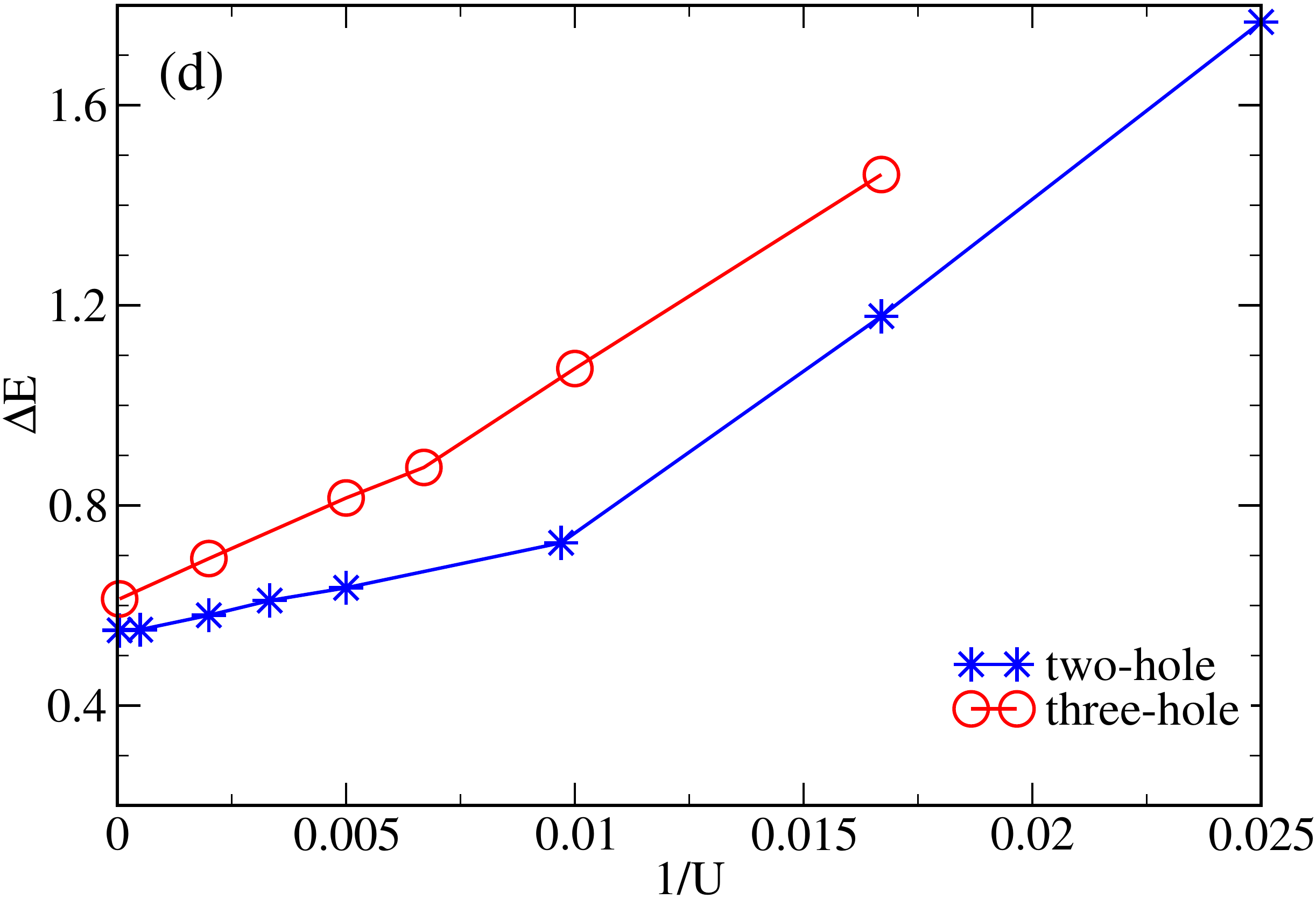}
\begin{flushleft}
\caption{\label{E(S)} $E(S)-E(S_{max})$ versus $S$ on 16- (a) and 18- (b) sites lattices with three holes, and the width of spin spectrum $\Delta E$ as the function of $1/U$ on 18- (c) and 20- (d) sites lattices, respectively. }
\end{flushleft}
\end{figure}

\begin{figure}[h!]
\centering
\includegraphics[width=0.3\textwidth]{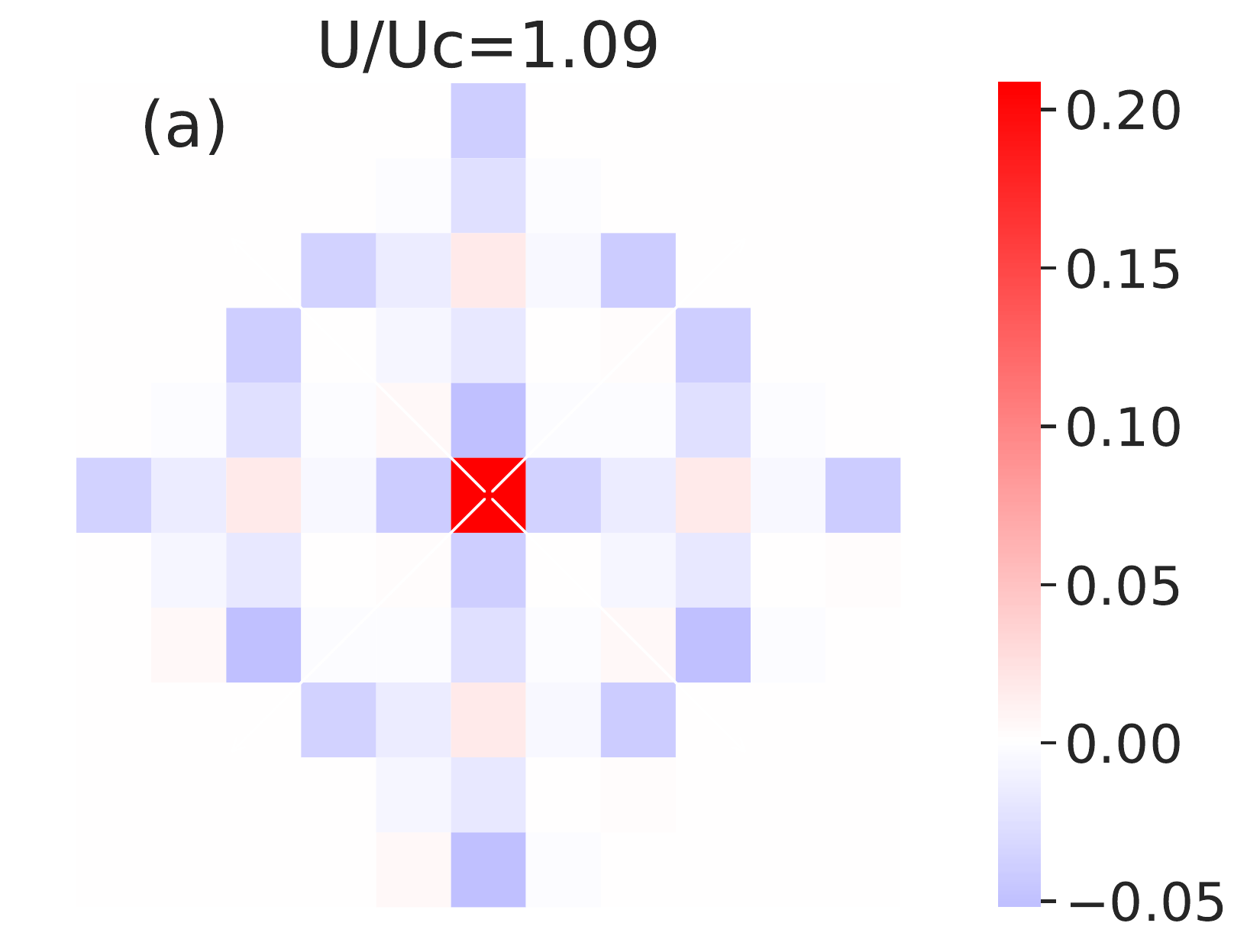}
\includegraphics[width=0.3\textwidth]{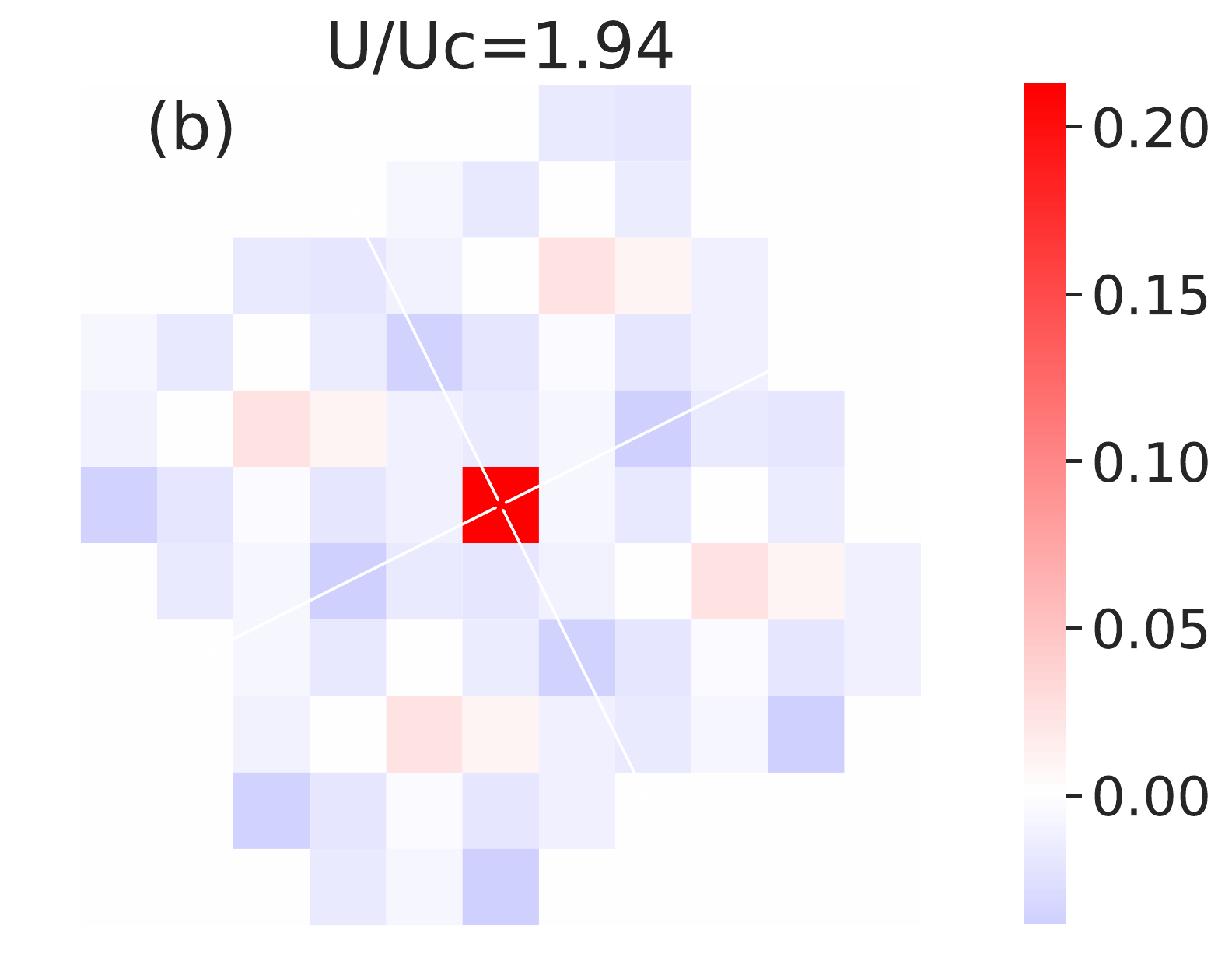}
\includegraphics[width=0.3\textwidth]{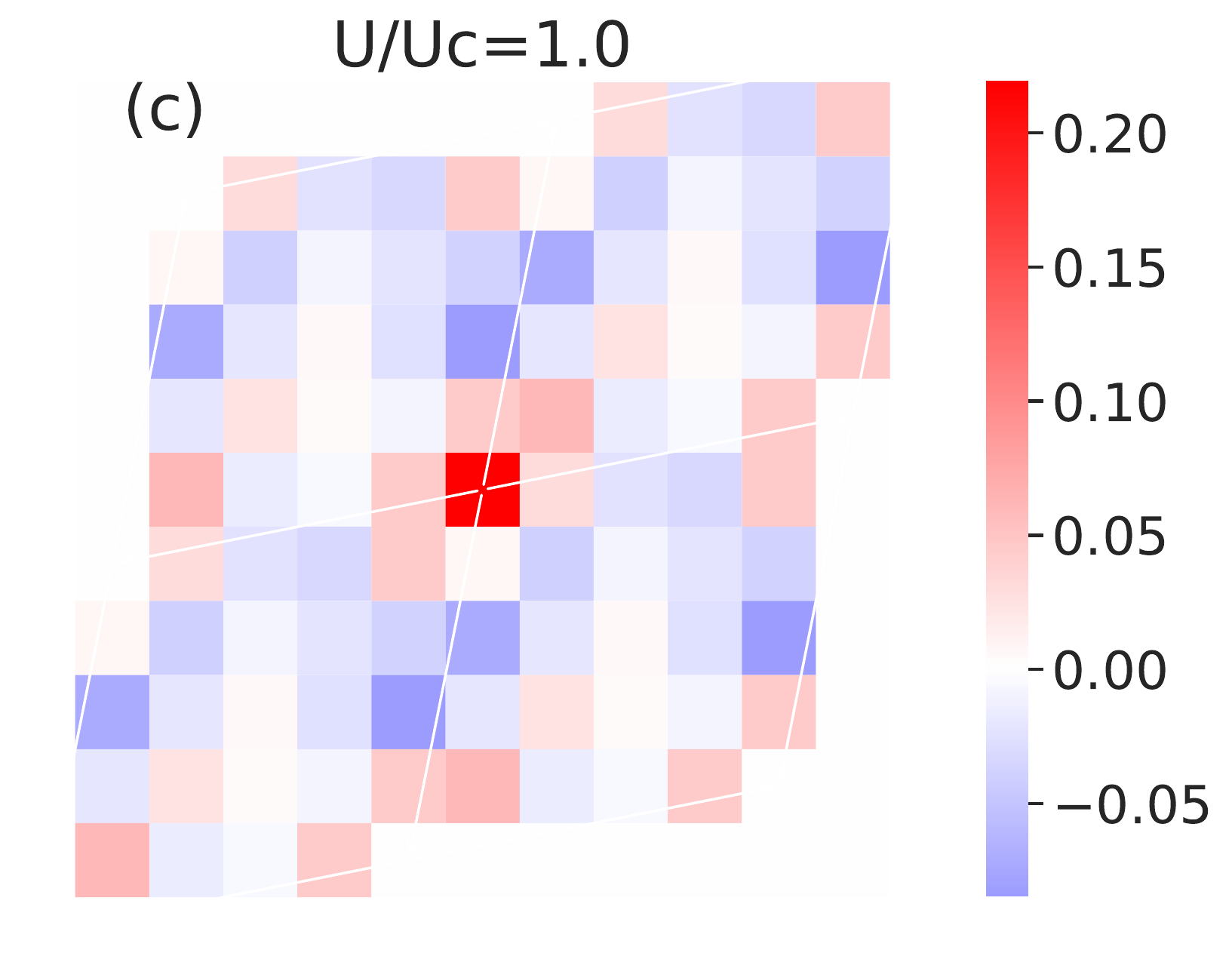}
\begin{flushleft}
\caption{\label{threeholes} The spatial spin distribution of the 18-, 20- and 24-sites lattices with 3 holes are provided in (a-c), $U_c$ are the same as those in two-hole system.} 
\end{flushleft}
\end{figure}

For three-hole systems,  we first study the possibility of partial spin-polarization on lattices larger than 16 sites. As been mentioned in the introduction, results of exact diagonalization (ED) based on an effective Hamiltonian~\cite{riera1989} show that the total spin of the ground-state on 8- and 16-sites lattices are $\frac{3}{2}$ and $\frac{7}{2}$ respectively. If this partial spin-polarization also exists for larger lattices, then for large enough $U$ there would be a phase of ferrimagnetism (a mixed type of anti-ferromagnetism and ferromagnetism). To investigate this problem, we have mainly calculated ground state energy as a function of total spin, $E(S)$, on 16- and 18-sites lattices.  

In Fig.~\ref{E(S)}(a), the ground state energy as a function of the total spin for the 16-site lattice is shown.
The results are presented for different $U$'s in comparison with the ED result for $U=\infty$~\cite{riera1989}, according to Eq.~\eqref{effHamil}. With increasing $U$, the $E(S)$ curves clearly converges to the ED result, and in fact the result at $U=20000$ already coincides with the ED result. These results show a partial spin polarization at $S=\frac 7 2$ in the large $U$ regime. 
However, on the 18-sites lattice, such a partial spin-polarization does not exist for any $U$. In 
Fig.~\ref{E(S)}(b), the results of $E(S)$ are plotted for different $U$'s for the 18-site lattice. The coincidence of the two curves at $U=20000$ and $U=50000$ indicates the convergence in the large $U$ limit. For any $U$  
the total spin of the ground-state always take the smallest value $S=0.5$ and thus there is no partial spin polarization. 
Similar results are also obtained on the 20-site lattice, where we have only performed calculations for $U=20000$. 
We therefore believe that the partial spin polarization result obtained in Ref.~\cite{riera1989} is only a finite size effect. 

To investigate the possibility of ferromagnetic domains in three-hole systems we have calculated the width of the 
spin spectrum $\Delta E=E_{max}-E_{min}$, where $E_{max}= \max(E(S)), E_{min}=\min (E(S))$ are the maximal and minimal values of the energy over all spin states for a given $U$. In our previous work~\cite{yun2021}, we find that for two holes systems  a clear change in the slope of $\Delta E(1/U)$ usually implies the formation of ferromagnetic domains.
The  width of spin spectrum $\Delta E$ for three holes systems on the 18- and 20-sites lattice are plotted in Fig.~\ref{E(S)}(c) and (d) respectively as functions of $1/U$, in comparison with those for two holes systems. 
For three-hole systems, the curve of $\Delta E(1/U)$ for the 18-sites lattice is simply a straight line,
while for the 20-sites lattice we see a small but clear change of slope. 
This indicates that on large enough lattices the three holes systems can also have ferromagnetic domains in the large $U$ regime. To support this, we have also investigated spatial spin distributions of three holes systems by calculating the spin-spin correlation functions.
The spatial spin distributions are presented in Fig.~\ref{threeholes}, where ferromagnetic domains are found on 20- and 24- site lattices. Comparing with the ferromagnetic domains in two-hole systems, the extent of ferromagnetic domains is smaller in this case.

\section{Conclusion}

The instability of Nagaoka ferromagnetism in the Hubbard model is studied for two and three holes systems on finite lattices. It is shown that the total spin of the ground-state takes the minimal value $S=S_{min}$, and there exists no partial spin polarization on lattices larger than 16 sites both on two- and three-hole systems. In the large $U$ regime ($U \geq U_c$, where $U_c$ is the critical $U$ of Nagaoka ferromagnetism in the one-hole system), we find the formation of ferromagnetic domains. Based on the analysis of binding energy and the statistical distance between two holes, we find a general feature of the ferromagnetic domain structure,  that holes tend to be far away from each other.

\begin{center}
{\large Acknowledgments }
\end{center}

Sujun Yun  would like to extend her sincere gratitude
to Nikolay Bogdanov and Giovanni Li Manni of the Max Planck Institute for Solid State Research, Youjin Deng in the University of Science and Technology of China and Qianghua Wang in Nanjing University for participating in discussion and providing support.
The authors gratefully acknowledge funding from the Max Planck Society.
Sujun Yun is supported by the national natural science foundation of China (Grant No.\,11805103 and No.\,11447204) and Jiangsu natural science foundation (Grant No.\,BK20190137).

\bibliography{article}

\end{document}


\title{Supplementary material: ferromagnetic domains of the Hubbard model by full configuration interaction quantum Monte Carlo methods}

\maketitle



\section{\label{app:spin-corr} Spin correlation functions}

Expressing the local spin operators as\cite{Paldus2012}
\begin{equation}\label{eq:local-spin-op}
	S_i^k = \sum_{\mu,\nu = \uparrow,\downarrow} = \sigma_{\mu,\nu}^{k} a_{i,\mu}^\dagger a_{i\nu}
\end{equation}
with the Pauli matrices\cite{Pauli1925}
\begin{equation}\label{eq:pauli-matrices}
	\sigma^x = 
	\begin{pmatrix}
		0	&	\phantom{-}1 \\	1 & \phantom{-}0 \\
	\end{pmatrix}, \quad 
	\sigma^y = 
	\begin{pmatrix}
		0 & -i \\ i & \phantom{-}0 \\
	\end{pmatrix}, \quad 
	\sigma^z = 
	\begin{pmatrix}
		1 & \phantom{-}0 \\ 0 & -1 \\
	\end{pmatrix}
\end{equation}
and the fermionic creation (annihilation) operators, $a_{i,\mu}^{(\dagger)}$ of electrons with spin $\mu$ in spatial orbital $i$,  results in the explicit expressions
\begin{align}
	S_i^x &= \frac{1}{2}\left( a_{i\u}^\dagger a_{i\d} + a_{i\d}^\dagger a_{i\u} \right), \nonumber \\
	S_i^y &= \frac{i}{2}\left( a_{i\d}^\dagger a_{i\u} - a_{i\u}^\dagger a_{i\d} \right), \nonumber \\
	S_i^z &= \frac{1}{2}\left( n_{i\u} - n_{i\d} \right),
\end{align}
where $n_{i\mu} = a_{i\mu}^\dagger a_{i\mu}$ is the fermionic number operator of orbital $i$ and spin $\mu$.

The (total) spin-spin correlation function, $\hat{\mathbf{S}}_i \cdot \hat{\mathbf{S}}_j$ is given 
\begin{equation}\label{eq:spin-corr-start}
	\hat{\mathbf{S}}_i \cdot \hat{\mathbf{S}}_j = \hat{{S}}_i^z \cdot \hat{{S}}_j^z + \hat{{S}}_i^x \cdot \hat{{S}}_j^x + \hat{{S}}_i^y \cdot \hat{{S}}_j^y
\end{equation}
with the individual terms as 
\begin{align}
	\hat{{S}}_i^z \cdot \hat{{S}}_j^z &= \frac{1}{4}\left(n_{i\u} - n_{i\d}  \right)\left(n_{j\u} - n_{j\d}  \right), \\
	\hat{{S}}_i^x \cdot \hat{{S}}_j^x &= \frac{1}{4}
	\left( a_{i\u}^\dagger a_{i\d} + a_{i\d}^\dagger a_{i\u} \right) 
	\left(a_{j\u}^\dagger a_{j\d} + a_{j\d}^\dagger a_{j\u} \right) \nonumber \\
	&= \frac{1}{4}\left( \cre{i\u} \ann{i\d} \cre{j\u} \ann{j\d}  + \cre{i\u} \ann{i\d} \cre{j\d} \ann{j\u} +  
	\cre{i\d} \ann{i\u}  \cre{j\u} \ann{j\d} + \cre{i\d} \ann{i\u} \cre{j\d} \ann{j\u} \right),\\
	\hat{{S}}_i^y \cdot \hat{{S}}_j^y &= \frac{1}{4}\left( \cre{i\d}\ann{i\u} - \cre{i\u} \ann{i\d}\right) 
	\left( \cre{j\d}\ann{j\u} - \cre{j\u}\ann{j\d}\right) \nonumber  \\
	&= \frac{1}{4}\left(\cre{i\d} \ann{i\u} \cre{j\d} \ann{j\u} - 
	\cre{i\d} \ann{i\u} \cre{j\u} \ann{j\d} - \cre{i\u} \ann{i\d} \cre{j\d} \ann{j\u} + 
	\cre{i\u} \ann{i\d} \cre{j\u} \ann{j\d}  \right).
\end{align}  

For singlets, $S = 0$, and $\braket{S^z} = 0$ \emph{ensembles}, defined as\cite{Kutzelnigg2010}
\begin{equation}\label{eq:s0-ensemble}
	\ket{S, \braket{S^z} = 0} = \frac{1}{\sqrt{2S + 1}}\sum_{m_s = -S}^S \ket{S,m_s}
\end{equation}
we have the additional symmetries in the RDM elements, valid for $\braket{S^z} = 0$ states, 
\begin{equation}\label{eq:syms}
	\Gamma^{i\u j\u}_{i\u j\u} = \Gamma^{i\d j\d}_{i\d j\d}, \qquad
	\Gamma^{i\u j\d}_{i\u j\d} = \Gamma^{i\d j\u}_{i\d j\u}, \qquad
	\Gamma^{i\u}_{j\u} = \Gamma^{i\d}_{j\d},
\end{equation}
with $\rho^{i\sigma}_{j\sigma} = \langle\Psi  \vert a_{i\sigma}^\dagger a_{j\sigma} \vert \Psi \rangle$, being the 1-RDM elements.  We can write  $\braket{S_i^z \cdot S_j^z}$ as
\begin{equation}
	\braket{S_i^z \cdot S_j^z} = \frac{1}{4}\left(\braket{\num{i\u}\num{j\u}} - \braket{\num{i\u}\num{j\d}} -
	\braket{\num{i\d}\num{j\u}} + \braket{\num{i\d}\num{j\d}} \right)
\end{equation}
and for $\braket{S^z} = 0$, with Eq.~\ref{eq:syms}, $\braket{\num{i\u}\num{j\u}} = \braket{\num{i\d}\num{j\d}}$ and $\braket{\num{i\u}\num{j\d}} =
\braket{\num{i\d}\num{j\u}}$\cite{Kutzelnigg2010}, as 
\begin{equation}
	\braket{S_i^z \cdot S_j^z} = \frac{1}{2}\left(\braket{\num{i\u}\num{j\u}} - \braket{\num{i\u}\num{j\d}} \right).
\end{equation}
For $i \neq j$, we can then identify
\begin{equation}\label{eq:as-rdms-1}
	\braket{\num{i\u}\num{j\u}}  = -\braket{\cre{i\u}\cre{j\u}\ann{i\u}\ann{j\u}} = \braket{\cre{i\u}\cre{j\u}\ann{j\u}\ann{i\u}} = \Gamma^{i\u,j\u}_{i\u,j\u}
\end{equation}
and similarly
\begin{equation}\label{eq:as-rdms-2}
	\braket{\num{i\u}\num{j\d}} = \Gamma^{i\u,j\d}_{i\u,j\d}.
\end{equation}
With Eq.\eqnref{eq:as-rdms-1} and Eq.\eqnref{eq:as-rdms-2} we can write $\braket{S_i^z \cdot S_j^z}$ as
\begin{equation}\label{eq:spin-corr-rdms-sds}
	\braket{S_i^z \cdot S_j^z} = \frac{1}{2} \left(\Gamma^{i\u,j\u}_{i\u,j\u} - \Gamma^{i\u,j\d}_{i\u,j\d} \right).
\end{equation}

We can also obtain the expectation value of the `full' $\braket{\mathbf{S}_i \cdot \mathbf{S}_j}$ spin-spin correlation function from a SD-based calculation. 
The full correlation function is given by
\begin{equation}\label{eq:temp-1}
	\mathbf{S}_i \cdot \mathbf{S}_j = \hat S_i^z \cdot \hat S_j^z + \hat S_i^y \cdot \hat S_j^y + 
	\hat S_i^x \cdot \hat S_j^x = \hat S_i^z \cdot \hat S_j^z - \frac{1}{2}\sum_\s \cre{i\s}\ann{j\s}\cre{j\bar{\s}}\ann{i\bar{\s}}.
\end{equation}

For $i \neq j$ the last term in Eq.\eqnref{eq:temp-1}, can be transformed to
\begin{equation}
	\cre{i\s}\ann{j\s}\cre{j\bar{\s}}\ann{i\bar{\s}} = \cre{i\s}\cre{j\bar{\s}}\ann{i\bar{\s}}\ann{j\s} = \Gamma^{i\s j\bar{\s}}_{j\s i\bar{\s}}.
\end{equation}
In total the expectation value $\braket{\mathbf{S}_i \cdot \mathbf{S}_j}$ can be obtained from RDMs via 
\begin{equation}
	\braket{\mathbf{S}_i \cdot \mathbf{S}_j} = \frac{1}{2}\left( \Gamma^{i\u j\u}_{i\u j\u} - \Gamma^{i\u j\d}_{i\u j\d} \right) - \Gamma^{i\u j \d}_{j\u i\d},
\end{equation}
due to the symmetry $\Gamma^{i\u j \d}_{j\u i\d} = \Gamma^{i\d j \u}_{j\d i\u}$.

For completeness,  we derive here the 
on-site, $i=j$, spin expectation value in terms of 1- and 2-RDM elements:
\begin{align}
	S_i^{z^2} &= \frac{1}{4}(\num{i\u} - \num{i\d})^2 = \frac{1}{4}[(\num{i\u})^2 + (\num{i\d})^2 - 2\num{i\u}\num{i\d}]\\
	\num{i\u}\num{i\d} &
 = \cre{i\u}\cre{i\d}\ann{i\d}\ann{i\u} = \Gamma^{i\u i\d}_{i\d i \u}\\
	\braket{S_i^{z^2}} &= \frac{1}{4}\left(\rho_{i\u i\u} + \rho_{i\d i\d} - 2\Gamma^{i\u i\d}_{i\d i \u} \right),
\end{align}
where $\rho_{i\s i\s}$ is a 1-RDM element. 

The total on-site spin expectation value is given by
\begin{align}
	S_i^2 &= S_i^{z^2} + \frac{1}{2} \sum_\s \cre{i\s} \ann{i \bar \s} \cre{i\bar \s} \ann{i\s} =  S_i^{z^2} +\frac{1}{2}\sum_\s \num{i\s}(1 - \num{i\bar \s})  \\
	&= \frac{1}{4}(\num{i\u} - \num{i\d})^2 + \frac{1}{2} (\num{i\u} + \num{i\d}) - \num{i\u} \num{i\d} \nonumber \\
	&= \frac{3}{4}\left( \num{i\u} + \num{i\d} - 2 \num{i\u}\num{i\d} \right) \nonumber \\
	\braket{S_i^2} &= \frac{3}{4}\left( \rho_{i\u i\u} + \rho_{i\d i\d} - 2\Gamma^{i\u i\d}_{i\d i \u} \right).
\end{align}




\bibliographystyle{apsrev4-2}
\bibliography{article}